\pdfoutput=1
\documentclass[11pt, oneside]{article}   	
\usepackage{geometry}                		
\usepackage{graphicx}				
\usepackage{amssymb}

\usepackage{multirow}

\title{Autonomics: an autonomous and intelligent economic platform and next generation money tool}

\author{Benjamin Munro and Julia McLachlan}

\date{March 23rd 2015}

\begin{document}

\graphicspath{{images}}

\newcommand*\vtick{\textsc{\char13}}

\maketitle

\newpage

\begin{center}

\begin{figure}
\centering
\includegraphics[scale=0.6]{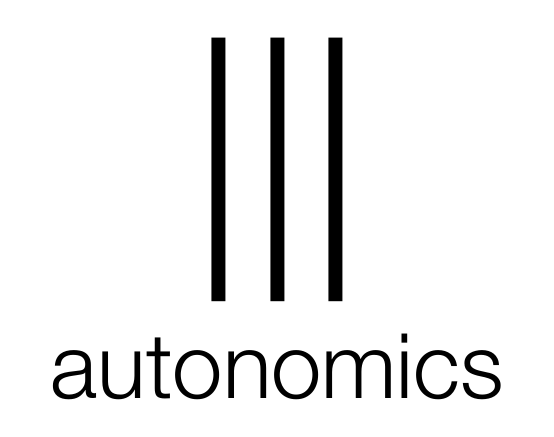}
\end{figure}

DESIGN SPECIFICATION FOR A NEXT GENERATION GLOBAL ECONOMY \\[10pt]

ONE -- Ontologically Networked Exchange \\[10pt]

Version 1.2\\[10pt]

THE HUMAN BLOCK CHAIN PROJECT -- A SUSTAINABLE ECONOMY \\[10pt]

A PROPOSAL FOR AN AUTOMATIC, ALGORITHMICALLY COMPUTATIONAL AND INTEGRATED CONSENSUS, TRUST AND DIGITAL MONEY ENGINE\\[10pt]

Comprising Proof-of-Autonomy (PoA) digital money, \lq mined\rq\ by human consensus and trust within an ontologically unified contract stack, and; an oracle machine (super recursive algorithm) for a trustable global artificial intelligence and Autonomy of Things. \\[40pt]

\emph{\lq\lq If we try to run the economy for the benefit of a single group or class, we shall injure or destroy all groups, including the members of the very class for whose benefit we have been trying to run it. We must run the economy for everybody.\rq\rq} \footnote{Henry Hazlitt \emph{\lq Economics in one lesson\rq}} \\[100pt]

CONTACT -- www.autonomics.io and contact@autonomics.io \\[20pt]

Copyright 2015 Benjamin Munro and Julia McLachlan. All rights reserved.

\end{center}

\newpage

\tableofcontents

\listoffigures

\newpage

\begin{abstract}	

We propose a high-level network architecture for an economic system that integrates money, governance and reputation. With this system we introduce a method for issuing, and redeeming a digital coin using a mechanism that solves large scale economic challenges and aims to create a sustainable global economy and a free market.

To maintain a currency\vtick s value over time, and therefore be money proper, we claim it must be issued by the buyer and backed for value by the seller, exchanging the products of labour, in a free market. We also claim that a free market and sustainable economy cannot be maintained using economically arbitrary creation and allocation of money.

Nakamoto\cite{nakamoto}, with Bitcoin, introduced a new technology called the cryptographic blockchain to operate a decentralised and distributed accounts ledger without the need for an untrusted third party. This blockchain technology creates and allocates new digital currency as a reward for \lq proof-of-work\rq, or \lq mining\rq, to secure the network.

However, no currency, digital or otherwise, has solved how to create and allocate money in an economically non-arbitrary way, or how to govern and trust a world-scale free enterprise money system.

We propose an \lq Ontologically Networked Exchange\rq\ (ONE) that is organised with purpose as its highest order domain. Each purpose is defined in a contract, and the entire economy of contracts is structured in a unified ontology. This unified ontology comprises just one of each contract type. As the mechanism issues money for labour transactions, governance is ensured by the economic imperative to collaborate in consensus, and in common purpose. Each transaction completed in the unified ontology of unique contract blocks is also unique, solving the double spending problem. We define this very many block chained network as the \lq blockcloud\rq. This forms the foundation of an economic operating system.

Unlike current blockchain platforms such as Bitcoin, whose networks are secured using economically arbitrary cryptographic proof-of-X (PoX) systems, we claim to secure the ONE network using	economically non-arbitrary methodologies and reasonable economically incented human behaviour. We define trust as transactions completed reliably over time in a given context or purpose.  By offering market participants contextualised transactional trust data, economic decisions are made based on reputation and reliability. Decisions influenced by trust and reputation help to secure the network without an untrusted third party. These properties combine in a unique way to solve the majority attack problem.

The stack of contracts, organised in a unified ontology, functions as a super recursive algorithm, with individual use programming the algorithm, acting as the \lq oracle\rq. The state of the algorithm becomes the \lq memory\rq\ of a scalable and trustable artificial intelligence (AI). This AI offers a new platform for what we call  the \lq Autonomy of Things (AoT)\rq. \\[10pt]

\noindent \emph{\lq\lq The art of progress is to preserve order amid change and to preserve change amid order.\rq\rq} \footnote{Alfred North Whitehead}

\end{abstract}

\newpage

\section{Introduction}	

Fiat currencies are based on confidence and inevitably return to their intrinsic value of zero. Rising prices are a symptom of an expanding currency supply, issued by capitalising non-value, and to non-producers thereby eroding market efficiency and destroying purchasing power. Humanity needs a money mechanism that accounts for a dynamic money supply and maintains purchasing power.

True wealth can be ensured, protected and sustained economically. Money is a trading tool that stores this wealth whereas currencies leak it away.

We propose a new digital money tool, \lq PoAcoin\rq\, which facilitates a truly free market and a sustainable economy. This new money tool functionally integrates money, governance and reputation to enable a globally unified accounting system. Unlike any economic model before in history, this model aims to provide the ideal mechanism for issuing real money into existence to capitalise those industrial endeavours that profit the common good. It motivates participants by agreement, transparency and wealth.

Currently the money tool and accounting system is being operated by governments and private banks. The problem with issuing money as credit is that interest on this debt increasingly drains value from the economy back to the issuers. The way banks create and allocate capital fails to meet free market needs. No currency to date, including gold, fiat or digital, has solved fundamental problems of creation, allocation, backing and redemption, or security of value.

Governments must issue money by spending, and inevitably will either spend on non-productive enterprise creating both inflation and the depreciation of money, or on productive enterprise -- and erode the competitive free market mechanism.

So, important questions are; how do we create and allocate money in a way that liberates both government and banking from a task which could be run by the free enterprise system itself? How do we reorganise human ecology, placing trust and consensus as having the highest value, and motivate all of humanity to be more transparent and collaborative, in projects that maximise profit and minimise plunder, nuisance and harm? And, how do we systemically end the externalisation of costs throughout the entire economy? \\

We define a free market as one where producers and consumers are free to trade voluntarily, and the market itself is responsible for the issuing and backing of money in a democratic process. Within the free market exchange mechanism, money is issued by the buyer and backed for value by the seller. The challenge is how to extend this two party exchange platform to a global scale.

Until now, the solution to the challenge of successfully integrating governance, money and trust hasn\vtick t been possible, however, with the evolution of the Internet and the emergence of block-chain type technologies, a world-scale economic mechanism is now possible and is proposed herein. \\

\emph{\lq\lq Orthodox economics has never yet been anything but the class economics of the owners of debts.\rq\rq} \cite[page 7]{soddy}

\subsection{Background}

In many ways, the current financial mechanism, which has evolved over hundreds of years, has been very successful in capitalising industry and  expanding the global marketplace. We\vtick ve seen the industrial age, the information age, and new technologies in every field of science that have been of immense benefit to humanity.

However, because banks create money as credit (debt), current banking practice has also lead to massive wealth disparity through charging interest on this debt, and, by increasingly allocating new money, as debt, to the holders of capital and assets to reduce bank lending risk, rather than to the productive working economy. \\

\emph{\lq\lq New money is principally created by commercial banks when they extend or create credit, either through making loans, including overdrafts, or buying existing assets.\rq\rq} \cite[page 6]{nef} \\

As we\vtick ve discussed, in addition to the failure of modern banking to create and allocate money in a way that creates a sustainable economy and a free market, over the course of history we\vtick ve built an artificially competitive and fragmented marketplace, with very many separate money, government and reputation systems. It seems almost impossible to redesign the whole system from the ground up. Even digital currencies, and associated blockchain technologies, of which many claim to be a solution of sorts, are subject to many of the same challenges. \\

\lq\lq \emph{How can a government, however well disposed, possibly provide money for private enterprise since, as we know, it cannot issue it without buying, and the more it buys the more it invades private enterprise and develops state ownership? ...  Why must our genius for increasing production be forever thwarted by man's inability to requisition his production into consumption?} \rq\rq \cite[page 28]{riegel} \\

With increasing tension between the benefits of expanding the marketplace, and the ill-fitted nature of incumbent mechanisms, it has become obvious our current systems are under duress. As the world looks to existing establishments for an answer to how to operate an increasingly globalised economy, we argue that is critically important, from a system design perspective, to re-think money at a fundamental level. How do we do this in a way that doesn't arbitrarily allocate power and that is governed in a transparent and trustable way? \\

\lq\lq \emph {Money power is the very essence of sovereignty and the failure by the citizen to assert it renders democracy futile.} \rq\rq \cite[page 30]{riegel}

\subsection{Problems}
Being functionally relevant to one another, money, governance and reputation systems must be integrated successfully to cause a democratic, free enterprise system. We cannot arbitrarily join disparate system parts together to create a functionally integrated whole. \\

\emph{\lq\lq We cannot solve our problems with the same thinking we used when we created them\rq\rq} \footnote{Albert Einstein} \\

And, therefore, our challenges are; how do we;

\begin{enumerate}
\item Build and operate a world-scale economic system that is stable and sustainable?
\item Intelligently create, allocate, back and redeem money within the free enterprise system itself?
\item Build an integrated system, so that the free market can adequately, democratically and scientifically both govern and trust itself?
\item Have all prices bear only, and all their costs, and organise production and consumption in a unified global system, to avoid externalising costs?
\end{enumerate}

The main problems with the current financial mechanism are:

\begin{enumerate}

\item \lq\lq It permits money to be issued privately, only by a limited number of persons and corporations who have bank credit, and makes such credit subject to fee. Thus it establishes credit as a privilege rather than a right, and makes it an object of profit rather than a utility to further the production and distribution of wealth. It denies to producers generally the right to issue money, thus making it impossible to expand buying power to potential producing power. This results in defeating the mass production system.

\item It permits the government to issue unbacked money. The only way the government could back its money issues would be to go into the production of goods and services; and this would compete with private business. Thus the problem offers the two horns of a dilemma, both of which lead to socialisation. If it backs its money issues with goods and services (and there is no other way it can be backed), it executes a frontal attack on private enterprise. If it issues money without backing it (as it is doing), it executes a flank attack on private business through inflation - since to issue money without creating equivalent values is to inflate.

\item It permits ambitious or designing or fanatical men who are in control of government to light the fires of war, threatening the lives and fortunes of untold millions. This terrible power lies solely in the political money system since armaments spring from money and money springs from government fiat, whereas it should spring only from the fiat of the people who would thus hold the veto power.\rq\rq \cite[page 9]{riegel}

\end{enumerate}

From the \emph{New Economics Foundation}, the main problems of the current system are:

\begin{enumerate}
\item \lq\lq Although possibly useful in other ways, capital adequacy requirements have not and do not constrain money creation and therefore do not necessarily serve to restrict the expansion of banks\vtick\ balance sheets in aggregate. In other words, they are mainly ineffective in preventing credit booms and their associated asset price bubbles.

\item In a world of imperfect information, credit is rationed by banks and the primary determinant of how much they lend is not interest rates, but confidence that the loan will be repaid and confidence in the liquidity and solvency of other banks and the system as a whole.

\item Banks decide where to allocate credit in the economy. The incentives that they face often lead them to favour lending against collateral, or existing assets, rather than lending for investment in production. As a result, new money is often more likely to be channelled into property and financial speculation than to small businesses and manufacturing, with associated profound economic consequences for society.

\item Fiscal policy does not in itself result in an expansion of the money supply. Indeed, in practice the Government has no direct involvement in the money creation and allocation process. This is little known but has an important impact on the effectiveness of fiscal policy and the role of the Government in the economy.\rq\rq \cite[page 7]{nef} \\
\end{enumerate}

According to Riegel, a sound money system must be separate from the political system, otherwise a corrupted profit motive will result. \\

\emph{\lq\lq We must first of all assume responsibility for the solution of our primary problem of prospering ourselves. To do this we must master money ... Government must be governed by a principle that defines the separate spheres of business and politics. When we take the money power out of politics, and allocate it to its natural sphere in private enterprise, we establish a proper coordination between the profit and non-profit motives of society. Without this allocation the two spheres are in constant conflict, breeding all manner of pressure groups and isms that seek to reconcile the irreconcilable. Money is an instrumentality of the profit motive and must be issued and backed only by private enterprisers.\rq\rq} \cite[page 7]{riegel} \\

The volume of money in an economy should be commensurate with productivity, rather than being controlled by private interests or politics. \\

\emph{\lq\lq Our research finds that the amount of money created by commercial banks is currently not actively determined by regulation, reserve ratios, the Government or the bank of England, but largely by the confidence of the banks at any particular period in time.\rq\rq} \cite[page 23]{nef} \\

By designing a money tool that inherently reflects and values the relationship between creditor and debtor, ONE is able to re-establish the proper role of money in an economy. \\

\emph{\lq\lq We have seen that the roots of an unstable money and banking system may lie in a mistaken understanding of money's nature by successive sovereigns and governments. ... The misunderstanding of money as a commodity rather than a creditor/debtor relationship has so dominated the imagination of our leaders and economists that they have allowed the development of a monetary policy regime that, despite repeated crises, remains steadfast.\rq\rq} \cite[page 141]{nef} \\

The private enterprise system itself could operate the money system. This allows production and consumption to operate effectively without being manipulated by non-market forces. \\

\emph{\lq\lq The very essence of the principle of private enterprise is the power to acquire and dispose of property. Since to acquire or dispose of property requires exchange and since the government or its creature, the banker, may veto exchange by withholding the exchange media, it can be seen that there is no private enterprise system in the full sense. Ours is, and has been from the beginning of political money, a political enterprise system, completely dominated by government directly or through its satellite, the banking system.\rq\rq} \cite[page 27]{riegel}

\subsubsection{Failing to create money debt-free}

\emph{\lq\lq Banks create \lq money\rq\ by book entries which are an equivalent of debt against the community. The specific function of the banking system is the creation of debt, or \lq\lq negative money.\rq\rq} \cite[page 104]{robertson} \\

Debt as money makes the current system perpetually dependant on \lq growth\rq.  How can a system that is dependent on perpetual, and not necessarily productive, \lq growth\rq\ create a sustainable economy? Perpetual growth and sustainability are fundamentally incompatible. \\

\emph{\lq\lq This debt is automatically self-cumulative and irredeemable.\rq\rq} \cite[page 185]{robertson} \\

\emph{\lq\lq Thus modern capitalism gave birth to a hierarchical form of regulatory control, with the central bank at the apex of the hierarchy ... The hope was that the central bank could control the quantity of commercial bank money ... however, deregulation and developments in technology have brought us to a situation where commercial banks now completely dominate the creation of credit and, hence, the money supply.\rq\rq} \cite[page 137-138]{nef}

\begin{figure}
\centering
\includegraphics[scale=0.6]{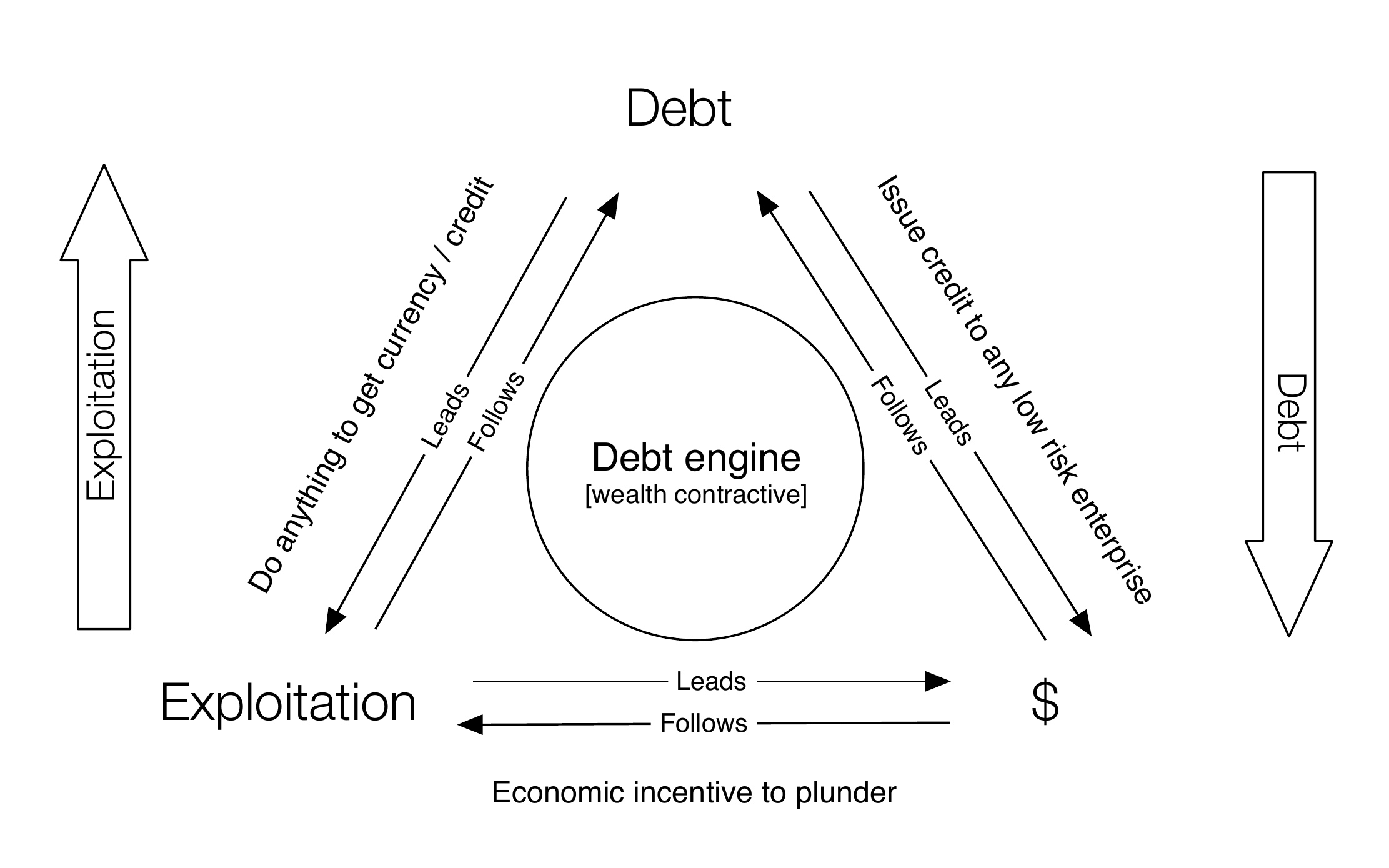}
\caption{The debt based money engine - arbitrary issuance and incentives}
\end{figure}

\subsubsection {Failing to allocate money to free market production}

\emph{\lq\lq Historical evidence suggests that left unregulated, banks will prefer to create credit for non-productive, financial or speculative credit, which often maximises short-term profits.\rq\rq} \cite[page 24]{nef} \\

The current money system, creates a wealth gap, increasingly favouring larger, wealthier, companies and asset holders. As the big get bigger, and the rich get richer, labour is increasingly forced to work for these ever bigger companies. \\

\emph{\lq\lq The political money system cannot help but operate adversely to the small enterpriser because his capital resources, on which banking credit is based, entitles him to such a small credit that the process of qualifying involves an expense far out of proportion to the possible interest income for the banker, and his profit possibilities are too limited to excite in the banker desire for participation. Therefore, probably 95\% of all enterprises are below the banking line. The usury laws are, in effect, laws against loaning to small enterprises because they confine loans to sums that are profitable at the maximum rate stipulated ... While the little fellows must engage in cut-throat competition, the big fellows have only a modified competition. To get from the nether to the upper level, the investment banker offers to smaller units the escape of amalgamation and thus he sits in on the profits and management of the new aristocratic corporation which thus acquires a competitively privileged position over the remaining nether group. Thus the political money system forces bigness as a means of survival. To be little is to be excluded from money power.\rq\rq} \cite[page 27]{riegel} \\

We need a money system that funds enterprise at any level, regardless of size. \\

\emph{\lq\lq Unless we find a method whereby each enterpriser shall be able to create exchange power commensurate with his size, we practice only sham competition and make a mockery of so-called free enterprise.\rq\rq} \cite[page 27]{riegel} \\

\subsubsection {Failing to back money's value}
Money must be backed for value to avoid inevitable debasement, and to form the basis of a sustainable economy. Without the economically non-arbitrary backing of a money\vtick s value, the money tool itself won\vtick t function over a long time, or be part of a functional accounting system. By the very nature of the money tool itself, money must be backed for value in a free market exchange, that is, by the seller. The current financial mechanism and its money tool, fiat currency, attempts to back value, but using economically arbitrary methodologies.

\subsubsection {Failing to redeem money}
Money forms the basis of a creditor/debtor relationship, as a promise to pay (later) for goods received now. Unless that promise to pay is eventually honoured, or redeemed, by the issuer, for other products in the marketplace, the promise becomes, inevitably, worthless. The fiat currency mechanism fails to redeem its money, and it therefore fails to operate as a proper money tool.

\subsubsection{Systemic risk}
This mechanism aims to mitigate the systemic risk inherent in the modern financial system.

\subsection{Mechanism}
\lq\lq \emph{Those who use mechanisms subserve the ends inherent in the mechanism.\rq\rq} \cite[page 13]{robertson} \\

The money mechanism must generate only beneficial aggregate effects for the whole group, when used by reasonable individuals optimising for their own good. This is the foundation of a sustainable economy. \\

\lq\lq \emph{...  if such a thing were done, little else in the way of arbitrary interference with and government control over the essential activities of men in the pursuit of their livelihood would be required. Indeed, just as now not one in a thousand understands why the existing money system has such power to hurt him, so, if it were corrected as here outlined, not one in a thousand would need to know or, indeed, would know, except by the consequences, either that it had been rectified or how it had been rectified.\rq\rq} \cite[page 3]{soddy} \\

We claim that;

\begin{enumerate}

\item The design of the economic system causes all subsequent effects that we refer to as our society and culture. \\

\lq\lq \emph{... it is the obsolete and dangerous monetary system that, primarily, is at fault.} \rq\rq \cite[page 1]{soddy}

\item A next-generation money tool is our most essential need.

We need a new money tool that is designed to systemically create very many, new and beneficial effects, both locally and globally. \\

\lq\lq \emph{Nothing useful can be done unless and until a scientific money system takes the place of the one now always breaking down.} \rq\rq \cite[page 3]{soddy} \\

\end{enumerate}

\subsection{Solution}

\emph{\lq\lq The new concept of money is that, to be sound and stable, and in adequate supply, it must spring solely from the same source from which all wealth springs, namely - the people, and that, to effectively coordinate with our mass production system, the people must issue the money necessary, to buy their production.\rq\rq} \cite[page 8]{riegel} \\

\emph{\lq\lq Either the people must have the money issuing power or the democratic power is lost. No power can transcend the political money power, once we accept its dominion, because money is a license to buy and a license to buy is a license to live. We are dependent upon money; and when any power outside ourselves controls money we are dependent upon that power.\rq\rq} \cite[page 44]{riegel} \\

The ONE economy provides the economic incentive, and the mechanism, to organise all labour in production to meet all market consumer demand. The mechanism itself creates, allocates, backs and redeems its own money, PoAcoin, in these free market transactions. \\

\emph{\lq\lq Indeed, a major source of objection to a free economy is precisely that it ... gives people what they want instead of what a particular group thinks they ought to want. Underlying most arguments against the free market is a lack of belief in freedom itself.\rq\rq} \footnote{Milton Friedman}

\section{Economics - the basics}

\subsection{Our economic premises}

\begin{enumerate}

\item The economy, society, community and culture are so interrelated as to be one thing.

\item Individuals naturally form groups for common purposes. By providing economic incentive to group in consensus we aim to organise the economy by common purpose as the highest order context or domain.

\item The specialisation of labour, of individuals and groups, creates economic production efficiency.

\item Individuals and groups will naturally relate economically to trade the products of their specialised labour.

\item The function of a marketplace is to exploit the efficacy of the specialisation of labour, and to provide an effective method for trading the products of labour. Without a marketplace we wouldn't be able to trade the products of our individual labour, and enjoy the wealth that this mechanism creates.

\item The function of money is to facilitate trade, and therefore the marketplace, by splitting barter so that any two parties can trade easily at any time. Money must be issued by the buyer and backed for value by the seller.

\item Money is a tool comprising an accounting system. The role of the accounting system is to expand the marketplace by standardising this issuance and redemption process using a common account ledger and a common money.

\item To maintain its value and ultimately its usefulness, the money instrument must be both issued and redeemed for value.

\item Money is issued for labour, and redeemed by the products of labour. This ensures money maintains high order value over long periods of time. Sustainable money value is key to a sustainable economy and a free market.

\item This market must be free - that is, operated by free people. Free people have rights -- to life, liberty and property, and the right to voluntary, informed, private, anonymous contract. For the two party exchange to honour human rights, and form the foundation of a democracy, it must be voluntary and voluntarily public. There must be a mechanism that economically incents transparency of transactions and rich trust data to build reputation.

\item The money tool is a mechanism, in which individual use causes aggregate group effects, these aggregate effects will be significant for a world-scale mechanism. We must build a money mechanism that maximises beneficial effects, and minimises harmful effects when used by many reasonable individuals. The mechanism must minimise the externalisation of costs to achieve this. This ensures that individuals acting independently and rationally according to each\vtick s self-interest maintain the best interests of the whole group. 

Prices must bear only, and all their costs. Costs are ultimately derived from labour.

\item We must have one integrated money system that operates as a single world-scale mechanism. This is essential to avoid externalising costs.

\item The money tool must functionally include governance and trust. That is, the tool itself must comprise an economy-wide governance mechanism and reputation system. Money, governance and trust must be integrated as they are functionally relevant to one another.

\item Scaling must be achieved without design or function compromise.

\item We aim to integrate money, governance and trust in a unified ontology of purposes.

\end{enumerate}

\subsection{Behavioural economics and incentives}

The desire for purchasing power is at least derived from the evolutionary imperative to be safe. Individual safety is assured through purchasing power, and group safety is assured by an economic mechanism that causes beneficial group effects and doesn\vtick t create the \lq tragedy of the commons\rq.  Purchasing power provides opportunity to participate in the marketplace of all products, to meet our individual needs. Purchasing power and the marketplace are our freedom to realise the value gained from the specialisation of labour, and collaboration. \\

It is assumed that a reasonable person will tend to follow basic economic incentives. These are;

\begin{enumerate}
\item Labour for money to increase purchasing power, in order to participate in the free market place.
\item Use this purchasing power to consume products of the marketplace.
\item Work more, and leisure less as purchasing power decreases and wages increase.
\item Work less, and leisure more as purchasing power increases and wages decrease.
\item Compete for wages  in the free market.
\item Be willing to pay higher prices for more trustable labour, and lower prices for less trustable labour.
\end{enumerate}

\subsection{Money is a tool}

Tools influence the behaviour of those that use them, particularly the money tool. Compared to fiat currency, a new money tool could be as different as an abacus and a super-computer.

\subsection{The function of money}

\lq\lq The four cardinal truths of money practice are: The Purpose of Money, The Source of Money, The Backing of Money, and The Democracy of Money.

\begin{description}
\item[The purpose of money] is to facilitate barter by splitting each transaction in halves, obviating the delivery of value by one trader (the buyer) and permitting the other trader (the seller) to make requisition for his half upon any trader at any time. This is the sole purpose of money. Any effort to employ it to influence prices or control trade is perversive.

\item[The source of money] is the trader (the buyer) who receives his half of the barter. Since it arises out of the buying process, and is based upon the evaluation of the acquired value made by the buyer, it is obvious that it can have no other source, and is created only by the act of paying for a purchase.

\item[The backing of money] Money is given its material backing by the seller through acceptance in exchange for value. Its moral backing is the buyer's pledge to accept it for equivalent value in free exchange.

\item[The democracy of money] Since trade is democratic, and since money is an instrument for facilitating trade, and since it can arise only from a trader in the act of buying, and be backed only by a trader in the act of selling, it is obvious that money is an instrument of democracy and the essence of man's sovereignty over business and government.\rq\rq \cite[page 56]{riegel}
\end{description}

\subsection{Specialisation of labour}

By incenting each individual with wages to contribute some specialised part of total production, a functional economy is able to organise all labour to produce all products for the marketplace. This \lq everyone doing their small part\rq\ is the foundation of wealth generation for the group. \\

\emph{\lq\lq It must be obvious to any thinking person that our progress from primitive to modern standards is due entirely to the specialisation of labour and that specialisation of labour implies the efficient producing of commodities that are not directly usable by the producer. This implies the necessity of facile exchange of products between producers, and that production can only be as profitable as exchange is facile. Therefore; whatever limits the facility of exchange limits the efficiency of production since production beyond the capacity of exchange is waste.\rq\rq} \cite[page 27]{riegel} \\

Optimally, money, as wages, should not only incent labour to work, but also incent labour to work on those aspects of production that generate wealth for the group, and not harm. \\

\emph{\lq\lq Man has learned that he can maintain a bare but precarious existence if he devotes his thought and labour to garnering or producing only those things that he consumes. To rise above this level of life he must become efficient in some occupation that produces exchangeable wealth. This specialisation of labour could yield no profit unless there be other men who likewise specialise; and it is further necessary that they meet to exchange their products. This implies a meeting or market place. Thus we see that three attitudes are basic to man's rise and continued progress, to wit: (a) the profit motive, (b) specialisation of labour to gratify the profit motive, and (c) exchange to realise the profit.} \cite[page 48]{riegel} \\

Individuals acting in the capacity of producers, earn wages, that then enables those individuals, acting in the capacity of consumers, to use their purchasing power to consume specific products of all labour. \\

\emph{The profit, or progress motive presses man toward means of greater production and he finds it in specialisation of labour. Greater production necessitates more exchange to realise the profit; and thus exchange becomes the neck of the bottle of production and consumption. Exchange, then, is the measure of human progress and limits or expands production because production (beyond subsistence) is purposeless without it. Therefore man can be only as wealthy as his exchange is facile.\rq\rq} \cite[page 48]{riegel} \\

\emph{\lq\lq Before you finish eating breakfast this morning, you\vtick ve depended on more than half the world.\rq\rq} \footnote{Martin Luther King}

\subsection{Money is an accounting system}

An accounting system requires a ledger that reliably credits and debits accounts, creating new balances. A common accounting ledger provides the backbone of an expanded marketplace. The expansion of the marketplace creates massive economic and social wealth through enabling the exchange of products derived from the specialisation of labour. \\

\emph{\lq\lq Money is the mathematics of value and is valueless in the same sense that mathematics is valueless. No amount of value can create money, but when men form a compact to trade with each other by means of accounting, in terms of a value unit, a money system is formed, and money springs into existence when any of them, by means of the act of paying for a purchase, incurs a debit in the accounting system. Conversely, money is destroyed by the process of selling in which a credit is earned against the previously incurred debit. Yet value is neither created nor destroyed by the process of creating and destroying money because money is but a concept.\rq\rq} \cite[page 48-49]{riegel} \\

The money instrument is the exchangeable token that is used to debit the buyers account and credit the sellers account. The money token would be of no use and no value without an accounting system. It\vtick s this ledger that enables a common money in an expanded marketplace. \\

\subsection{Freedom and a free market}

ONE is designed to allow each user the ability to establish and maintain voluntary privacy, ownership and control of their own data, accounts and contracts.

Freedom of the individual and the marketplace must be sustainable. So, we must design an economy that avoids depletion of some common resource when individuals act independently and rationally according to each\vtick s self-interest, otherwise we will become decreasingly free without understanding why.

Without a mechanism that includes effective group consensus, and collaboration, we can't have individual freedom.

\subsection{Externalising costs - the difference between profit and plunder}

It is natural that in creating value, making a profit will incur costs. If however, the total costs are greater than the total value created, the result is a loss, not a profit. When costs are externalised, by having someone, or something else bear the costs, then it is likely that, the so called profit, from the perspective of a small special interest group in the short term, is in fact a loss, for the whole group in the long term. When the costs are born by other parties without their consent or knowledge, the parties are coerced, and the result is plunder, and harm.

A sustainable economy must ensure, that when individuals acting independently and rationally according to each's self-interest they don't, in aggregate, behave contrary to the best interests of the whole group. \\

\emph{\lq\lq Profit is any increase in happiness acquired by moral means.\rq\rq\ \emph{and}  \lq\lq Morally acting man seeks to profit; immorally acting man seeks to plunder.\rq\rq} \cite{galambos} \\

However, if the economic mechanism that we all use systemically externalises costs, and incents its users to externalise costs, then the system itself is acting to overall plunder, regardless of the morality of its users, who may seem to be individually, albeit temporarily, profiting. The only way to solve this, is to have one global economic accounting system, that uses a unified ontology to account for everything. This is successful when no transaction coerces another party. \\

\emph{\lq\lq Rather, I am attacking an \emph{idea} which I believe to be false; a system which appears to me to be unjust; an injustice so independent of personal intentions that each of us profits from it without wishing to do so, and suffers from it without knowing the cause of the suffering.\rq\rq} \cite[page 18]{bastiat} \\

\emph{\lq\lq This is because all men seek happiness, and, when they achieve it, without coercing anyone else, they will find they have gained a profit.\rq\rq} \cite{galambos}

\subsection{Regulation - what to regulate and how}

A free market enterprise system requires that all regulation is provided by, and within, the mechanism itself. To eliminate potential bias or corruption, any, and all, economically arbitrary external influence must be avoided by design. However, to achieve this, the money mechanism itself must be incapable of harm. \\
 
\emph{\lq\lq If we can coordinate consumption with production we may develop our mass production to the point where the fullness of production will itself bring about the diminishment of hours of labour, the abolishment of child labour and the labour of the aged, and give us less work and more leisure, until the ideal balance between work and leisure is attained.\rq\rq} \cite[page 57]{riegel}

\subsection{Labour, wages, prices - purchasing power}

As purchasing power is ultimately the relationship between wages and prices, it is imperative that wage prices and product prices are related in a economically non-arbitrary way. ONE ensures that wages are commensurate with prices and vice versa.

If money is allocated to non-productive participants, normal wealth distribution will artificially skew. As wealth artificially re-distributes, so does relative purchasing power.

This money system puts an end to artificial, relative wealth, as a result of arbitrary money allocation which skews money distribution, and restores real general wealth of the common good, distributed normally.

\subsubsection{Unemployment and leisure}

Unemployment isn\vtick t necessarily a \lq bad\rq\ thing. What matters is that consumers have enough purchasing power to buy the products of all labour, and that their needs are efficiently met. In a very efficient economy, that maximally utilises the specialisation of labour, and maintains the right amount of high quality \footnote{see later section} money, lower levels of unemployment, and more leisure, are actually desirable.

\subsubsection{Jobs and the economic incentive to work}

One of the core requirements of a functional economy, is to provide sufficient economic incentives, wages, to labour, in order to encourage people to participate in production for the market's benefit. Participation, contribution and collaboration should be fundamentally incented for.

\section{The evolution of the money tool}
The three aspects we discuss in the context of the money tool\vtick s evolution are;

\begin{enumerate}
\item Its creation, allocation, backing and redemption
\item Its governance and regulation
\item Its accounts ledger
\end{enumerate}

Here we compare these three aspects changing with time over 4 distinct money types - which we call Money 1 through 4.

\subsection{Money 1.0}

\paragraph{Commodity money, I.O.U's, scarce metals}
Without common money or ledger of accounts, these small private and fragmented marketplaces, allowed local trade based on local reputation. However, these systems were limited geographically, and subject to inherent instability. These systems were unable to fund large scale enterprise and the expansion of industry and the marketplace.

\subsection{Money 2.0}

\paragraph{Fiat currencies, central and commercial banking}
With well organised private, centralised money, the state-and-bank-operated accounting system enabled the expansion of industry and the marketplace, with a common ledger and a common money. However, the system still suffers from arbitrary creation and allocation of credit as unbacked debt, and arbitrary political regulation. These systems have proven to be unstable and unsustainable.

\subsection{Money 3.0}

\paragraph{Crypto-currencies}
With a paradigm shift in decentralised ledger technology, cryptocurrencies have proven the possibility of a public, transparent accounting system without the need for an untrusted third party. However, these systems suffer from economically arbitrary creation, allocation, backing and redemption of currency though Proof-of-X blockchain rewards, and are still subject to the political regulatory environment. These crypto-currency systems are still entirely divorced from large scale economic concerns and how they capitalise industry. Many digital currencies are still subject to the Money 2.0 domain regulatory authority.

\subsection{Money 4.0}

\paragraph{Autonomics and PoAcoin}
Proof-of-Autonomy (PoA) coin, is a digital money, \lq mined\rq\ by human consensus and trust within an ontologically unified contract stack of common purposes. With a fundamentally integrated system, PoAcoin is able to create, allocate, back and redeem money within the free enterprise system itself.

The ONE economy is self-regulating because its functions are centralised, decentralised and distributed by contextualisation.

\section{The model - Autonomics and the blockcloud}
We define Autonomics as the science of a self-determining group. And we define the structure on which the system operates as the blockcloud.

The ONE mechanism has potential to organise the global free market to generate \lq broad and deep\rq\ wealth, and the more people that use it, the more it will realise its potential. From the ONE Engine point of view, it is buying the products of industry to increase the common good, as agreed on in consensus. It has unlimited money to do this.

\subsection{The objectives}

Our goal is to, with as few assumptions as possible, design a mechanism that will create all necessary effects for a next-generation global economy. We claim that under reasonable human use using economic incentives, this mechanism logically causes a free market and sustainable economy. \\

\noindent More specifically, we aim to:

\subsubsection{Create a free and voluntary market}

\emph{\lq\lq The True Democracy, then, is the free, unhampered market economy.\rq\rq} \cite[page 43]{galambos} \\

The only way to have a free market, is to fundamentally integrate money, governance and reputation, so that the free enterprise system can be self-determining.

\subsubsection{Internalise all costs and benefits}

The ONE mechanism aims to bring external costs to as close to zero as possible. \\

\emph{\lq\lq In economics, an externality is the cost or benefit that affects a party who did not choose to incur that cost or benefit.\rq\rq} \footnote{Wikipedia \lq\ Externality\rq}

\subsubsection{Create an economy that is efficient and sustainable}

Our aim is to motivate labour, with economic incentives, while mechanistically benefiting the group without harm, to optimise for profit of the common good; and to, provide a mechanism by which the sound purpose, issuance, backing and democratic nature of money can be scaled globally without corruption of the money tool itself.

\subsubsection{Secure money's long-term value}
PoAcoin\vtick s value is inherently secured by directly representing how the economy values working in consensus and trust. PoAcoin would not be debased. Its value is backed by its exchange for labour, and the products of labour. It could only be debased by the majority of working, collaborating people debasing their own worth.

\subsubsection{Maintain one global accounting system}

ONE works the same way everywhere. That is, its fundamental function scales iteratively and reliably across all contexts, including the geographical context. Consistency of function maintains the \lq level playing field\rq\ required for an effective and \lq fair\rq\ system. \\

\emph{\lq\lq If we try to run the economy for the benefit of a single group or class, we shall injure or destroy all groups, including the members of the very class for whose benefit we have been trying to run it. We must run the economy for everybody.\rq\rq} \footnote{Henry Hazlitt \emph{\lq Economics in one lesson\rq}} \\

\emph{\lq\lq For many things that seem to be true when we concentrate on a single economic group are seen to be illusions when the interests of everyone, as consumer no less than as producer, are considered. To see the problem as a whole, and not in fragments: that is the goal of economic science.\rq\rq} \cite[page 183]{hazlitt} \\

\emph{\lq\lq Economics, as we have now seen again and again, is a science of recognising secondary consequences. It is also a science of seeing general consequences. It is the science of tracing the effects of some proposed or existing policy not only on some special interest in the short run, but on the general interest in the long run.\rq\rq} \cite[page 175]{hazlitt} \\

and, in the context of a sustainable economy, we would to expect to see:

\begin{enumerate}

\item wages raised to the highest possible level,
\item sufficient purchasing power to meet needs while effectively balancing labour and leisure for everyone,
\item steady price levels and no inflation or deflation,
\item decreased bureaucracy and centralisation of government,
\item the assurance of real freedom, prosperity and democracy,
\item the preservation of peace, and to prevent injustice.

\end{enumerate}

\subsection{Knowledge}

\subsubsection{Public good, private good, and the common good}
It is up to the public to decide what is good for the public. However, to be effective, the public must operate within a mechanism that provides:

\begin{enumerate}
\item Trust data so that everyone has the ability to measure and see what is happening everywhere globally,
\item Consensus that offers fair and contextualised representation for everyone and 
\item Money that capitalises industrial activities, and makes producers and consumers wealthy.
\end{enumerate}

Without the right organisational mechanism, the common good is limited to libraries, hospitals and roads etc. However, the ONE engine framework is able to contextualise all contracts and transactions (issuance), so that the common good can increasingly, and intelligently, include specific aspects of people\vtick s individual lives. In other words, this mechanism can increasingly fund individual good as well, because we\vtick re able to define the common good as highly contextualised aggregates of all people\vtick s individual good. In this way, it enables all individuals to \lq compute\rq\ the common good as a function of individual good. 

For the purpose of this paper, these terms can be replaced with the idea of collaborative profit of a highly contextualised common purpose in which all people voluntarily participate.

\subsubsection{Centralised, decentralised and distributed}

Centralisation is not \lq bad\rq, and decentralisation \lq good\rq. It\vtick s important to distinguish what is centralised, decentralised or distributed and by what mechanism, and in what context.

For example; in the ONE economy, individuals can centralise by purpose in a contracting group, while different contracts are distributed contextually throughout the economy. Governance is centralised within each group, however, it is decentralised by context throughout all groups.

\subsubsection{Power and force}

In physics, power is the rate of doing work. It is equivalent to an amount of energy consumed per unit time. In ONE, work is defined in transactions, which are contextualised and centralised by purpose defined in contracts. Purpose is the reason for which something is done or created or for which something exists.

In physics, a force is any external effort that causes an object to undergo a certain change, concerning its movement, direction or construction. Force, in law, is unlawful violence threatened or committed against persons or property.

Power should not be able to force. It must be decentralised without losing its ability to be effective in its own context. Therefore, ONE decentralises power and distributes it throughout the economy by context, so that power remains effective within each group. The unified ontology of contextualised groups are able to collaborate via the mechanism. \\

Having centralised purpose (in a group) will allow for an increase in power far greater than the sum of its (individual) parts. It is \emph{how} we connect that makes us great.

Ultimately, it will be economic incentives that will motivate for common purpose, successfully out-competing any nuisance force.

\subsubsection{Work and trust}
Work is defined as any transaction that occurs in consensus within the ONE Engine contract stack. Completed transactions, work, is rewarded with PoAcoin, as a credit. Work can be completed by individuals, groups, robots and autonomous agents.

For the purpose of this mechanism, it is not enough for trust to be based on a feeling, opinion, guessing, religion or myth. Trust must be known, and specifically contextualised, not generalised or abstracted. Therefore for the purposes of the ONE system trust is defined and measured as transactions completed reliably in context (contract) over time. It is specific, measured data organised contextually in a unified ontology of purposes. The more any participant (human, robot or autonomous agent) transacts transparently, the more contextualised trust will be generated. This naturally, and intelligently resolves as an integrated reputation system.

The system offers reliable, system-wide, fully-contextualised trust data so that the free market is able to contract quickly and efficiently. Trust becomes a valuable aspect, because the free market will pay more for it and economically incents all labour to be effective, transparent, honest and reliable. 

Trust is the primary influencer of wage rates, and therefore economically valuable. Labour will be incented  by trust and for money. Trust\vtick s value is agreed on in the free market, setting wage rates, within every context. The market will aggregate around, and value trust. Therefore the market will aggregate around and value PoAcoin, and increasingly, securing its value.

By economically incenting trustable behaviour we allow participants to earn their way into trust and reputation, motivated financially within the free market. This is the right mechanism to motivate for trustable behaviour, and therefore build the foundation of an \lq ethical\rq\ economy and an economically restorative justice system.

\begin{figure}
\centering
\includegraphics[scale=0.6]{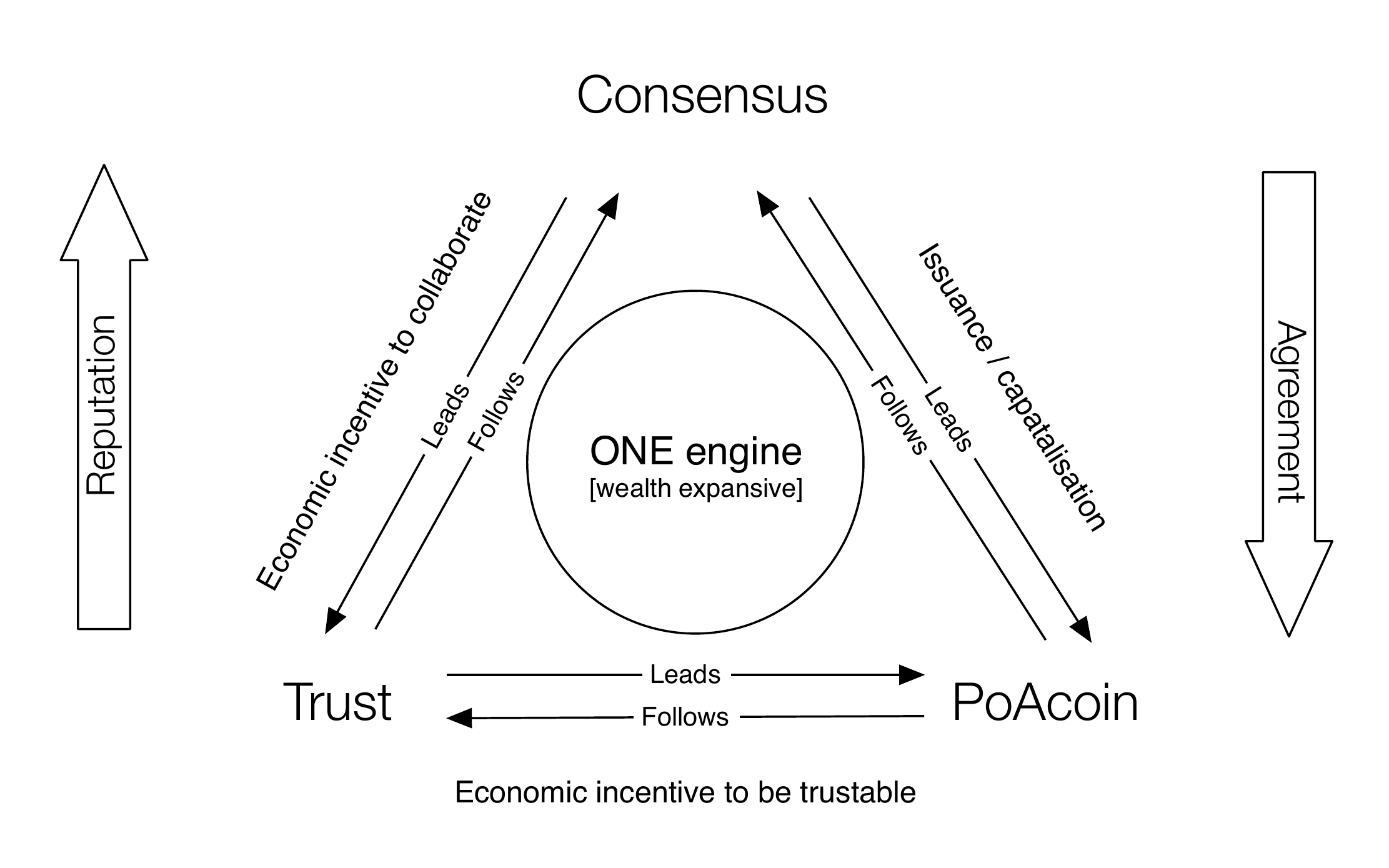}
\caption{The ONE Engine - integrating money, governance and reputation}
\end{figure}

\subsection{Outline of model}
ONE acts as an impartial, mechanistic intermediary, or clearing system, that organises production (labour) and consumption (products) contextually in a unified ontology. Producers and consumers are the same set of all people. An individual acting in the capacity of specific producer, labours to earn purchasing power and is paid PoAcoin, a claim on general products. When acting in the capacity of consumer the individual redeems their claim for specific products of general labour. This ensures that money maintains its value over time. \\

Proof of Autonomy is an aggregate of many \lq proofs\rq, which comprise work (as labour), stake (as reputation), and redemption (as consuming) in trust and consensus. This, overall,  reflects the degree of common purpose.

ONE, essentially, gets labour production for nothing when it issues PoAcoin as money claims. ONE then, technically, owes all labour, all the products that labour has produced.  then redeems PoAcoin for products it returns back to the marketplace.

Labour, and production is held in trust by the ONE system, and exchanged for PoAcoin in the market place. The PoAcoin received, or redeemed, is then used to redeem wage costs throughout the economy.

\subsection{Architecture and properties}

\subsubsection{Nodal tree, contract stacks, chains and transactions}
Contracts define purposes that fulfil and define higher order contracts (HOCs), and comprise and define lower order contracts (LOCs). That is; HOCs comprise and define LOCs. LOCs aggregate to fulfil and define HOCs. Of course these are relative terms. The HOC is an LOC relative to its own HOC. The LOCs are HOCs relative to their own LOCs.

This structure naturally supports the principle of the specialisation of labour. That is; LOCs are specialised contracts that aggregate to fulfil HOCs, or specifically, LOCs are specialised contracts fulfilling relatively general HOC contracts.

\begin{figure}
\centering
\includegraphics[scale=0.6]{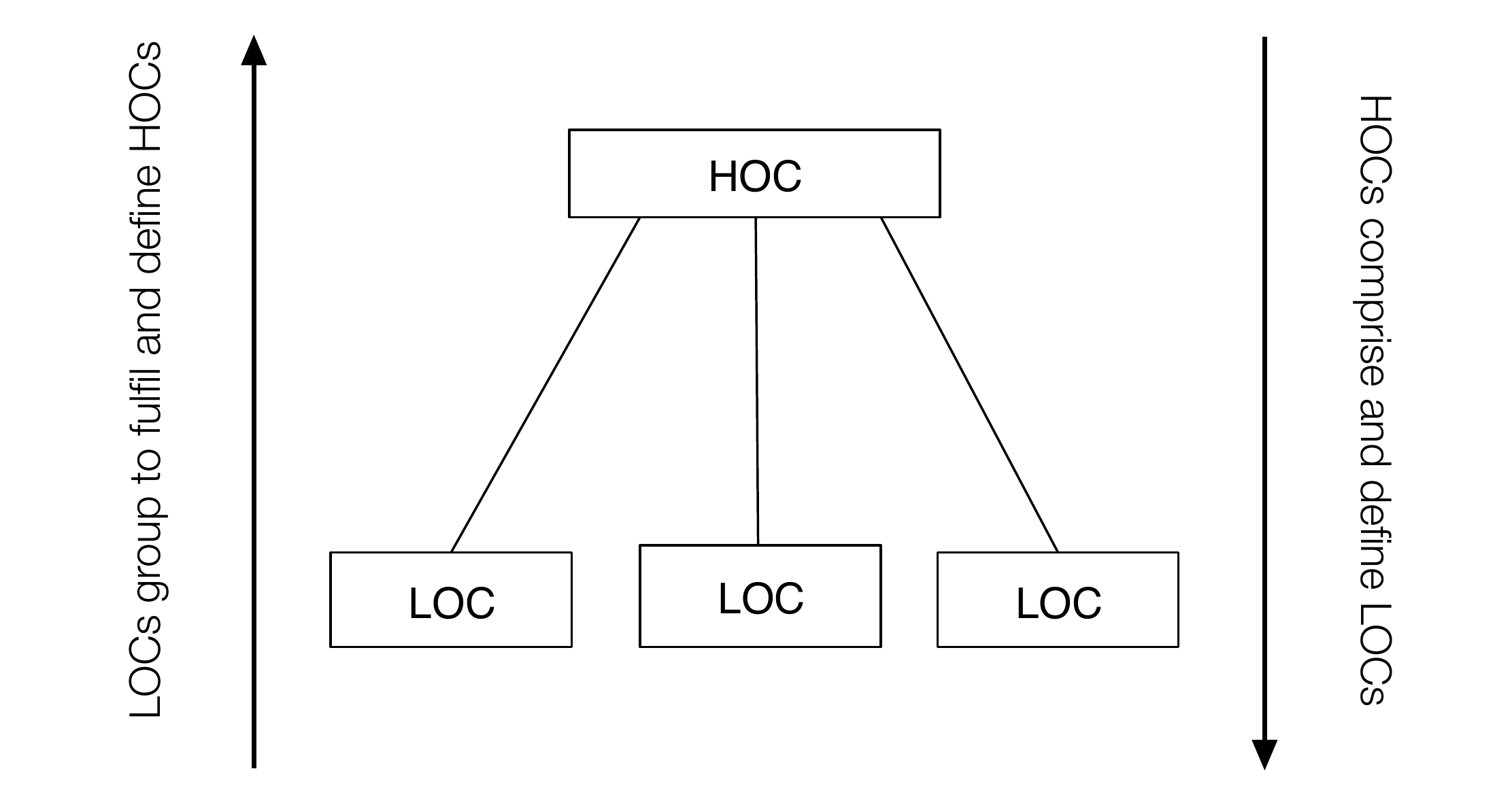}
\caption{Higher order, and lower order contracts}
\end{figure}

This contract stack structure organised in a unified ontology forces contract commonality, and therefore contextualised consensus.

All transactions occur within, and are defined by contracts.

This iterative structure forms the contract \lq stack\rq\ or contract \lq state tree.\rq

This takes the traditional producer-consumer model and uses an economy-wide recursive model.

The term \lq chain\rq\ describes a lineage of contract fulfilment, or a chain path of descending stacked contracts and component contracts.

\begin{figure}
\centering
\includegraphics[scale=0.7]{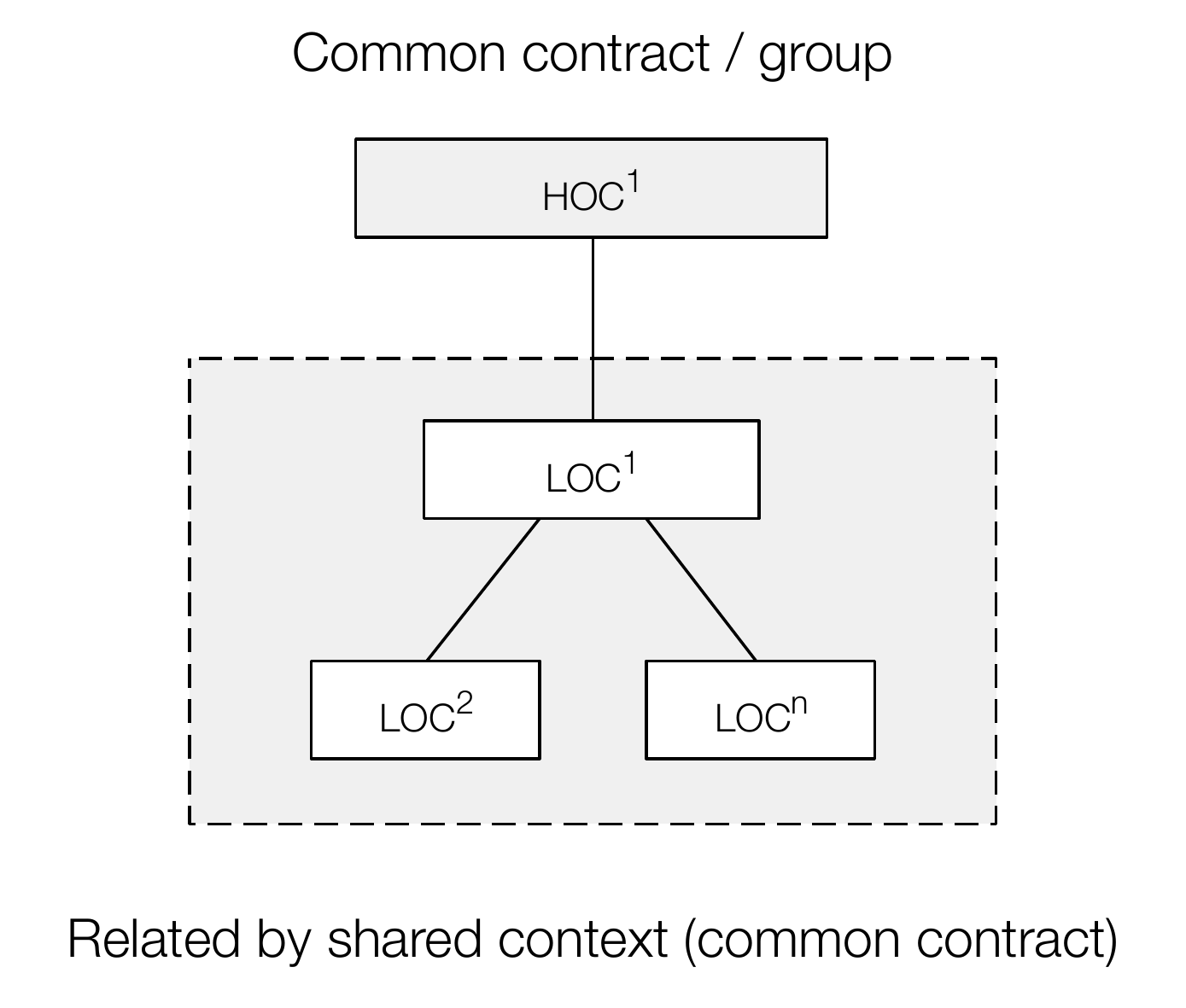}
\caption{LOCs share a common context in the HOC}
\end{figure}

\subsubsection{Groups}
In the ONE economy, the idea of companies and corporations is replaced with dynamic groups organising in common purpose.

\begin{figure}
\centering
\includegraphics[scale=0.8]{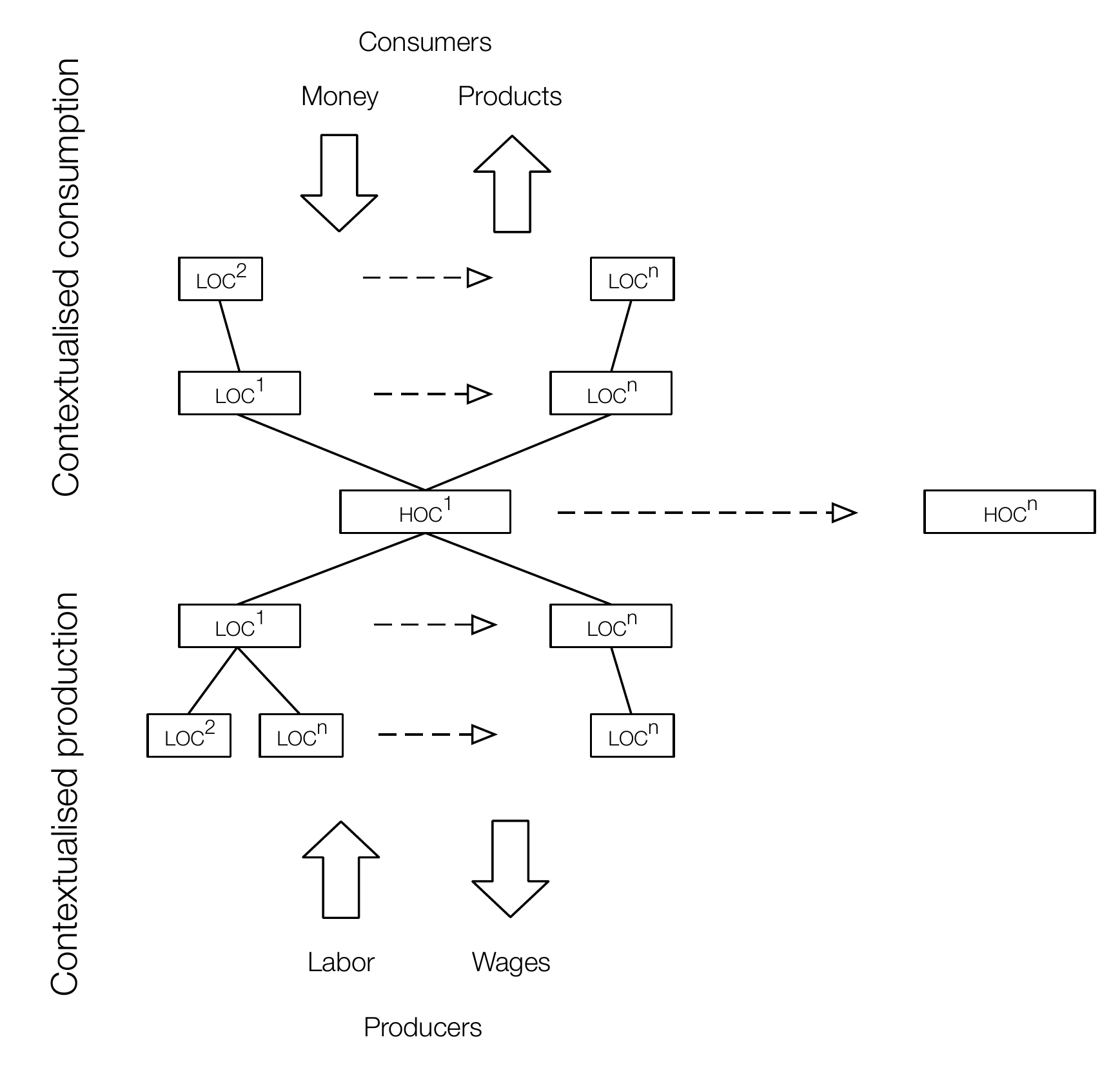}
\caption{The Autonomic Economy - Ontologically Networked Exchange (ONE)}
\end{figure}

Contracts are inevitably complex, yet mechanically simple, natural subsets of each other. All contracts (and therefore groups) stack for completeness (supersets-subsets), any incompleteness is represented automatically as a call-put in the marketplace. An expanded contract calling for completeness from LOCs, and an expanded contract \lq putting\rq\ to an HOC, these would be accepted in relevant consensus.

Any incompleteness indicates a call and/or put for contract changes. All contracts are stacked and aggregated throughout the entire nodal tree. So, looking \lq down\rq\ the nodal tree, towards any sub-groups (LOCs), one can \lq see\rq\ any incompletion in contract fulfilment - that is, if the purpose of the group is being fulfilled or not will naturally \lq call\rq\ for support in that field, effectively being the new \lq job\rq\ market. Looking \lq up\rq\ the nodal tree, into a super-group (HOCs), one can \lq see\rq\ what needs doing, and offer labour, for contract fulfilment. Dynamic reporting and custom \lq views\rq\ onto the dataset become trivially easy. This creates a dynamic requirements-fulfilment mechanism.

As long as the contract stack exists in completeness, or reveals puts and calls where it is incomplete, the ONE Engine ecosystem is a computational economic engine that capitalises industry.

In this dynamic system contracts, groups, are being created, modified and dissolved constantly.

Transparency allows for trust, which provides context for meaningful consensus decisions. Projects which humanity agrees are valuable to the common good are fully funded.

\subsubsection{The ONE Engine}

ONE provides the public a useful tool to allow all people to organise by consensus to profit the common good, and to have trust data to inform transaction decisions. By capitalising industry that achieves this, an entire global economy can efficiently satisfy the needs of the common good.

From the ONE Engine point of view, it is buying the products of industry for the global public to increase the common good, as agreed on in consensus. Money is issued debt free, without interest. The ONE engine has unlimited money to do this.

There\vtick s a time and space aspect to each transaction chain. The space aspect is the path through the consensus tree state, and each transaction is time stamped, in the time aspect. The state of any aspect changes constantly both in time and space through use.

The time aspect is synonymous with trust. That is, trust data builds over time in every aspect, deriving historical records in the blockcloud time domain.  PoAcoin is part trust in time, and part consensus in space, time stamped unique transactions in a consensus tree ontology. Trust and consensus are integrated aspects of PoAcoin.

\subsubsection{The double spending problem}

\emph{\lq\lq First of all, let us understand the problem. The purpose of bitcoin mining is to create a decentralised time stamping system, using what is essentially a majority vote mechanism to determine in which order certain transactions came as a way of solving the double-spending problem.\rq\rq} \footnote{Vitalik Buterin on \lq mining\rq, from the \emph{\lq Ethereum Whitepaper\rq}} \\

To prevent debasement of PoAcoin, through counterfeiting, we must ensure that double issuance, when purchasing labour, and double redemption, when purchasing goods and services is mechanistically impossible. The only way to ensure that each coin is unique, is to ensure that each coin issues or is redeemed through a unique transaction chain. We ensure transaction chain uniqueness by structuring all transactions, and all contracts in a unified ontology. This way, there is just one of each contract type, and any transaction within this structure is therefore a unique chain. Within this very-many-chained blockcloud, each context forms its own \lq blockchain\rq\ ensuring each transaction is unique, solving the double spending problem.

\subsubsection{The unified ontology}
The unified ontology comprises just one of each contract type, optimally exploiting the principle of specialisation of labour, which fulfils one of the requirements of a free market economy. The contract stack dynamic forces contracts into aggregate commonality, solving for consensus governance.

By operating within a unified ontology, the contract stack will organise free market labour in common purpose, driving consensus and issuing labour money for completed transactions. The system is \lq programmed\rq\ dynamically, and constantly through human economic use.

It\vtick s the contextualisation of consensus, trust and money issuance/redemption, maintained in an unified ontology that maintains the blockcloud in purposeful state.

This ontologically unified contract stack causes natural structure around purpose, rather than arbitrary group formation. It re-defines, entirely, previous ideas of companies, monopoly and competition.

Rather than have large numbers of similar contracts repeatedly used throughout the ecosystem, contracts would be constructed from standardised component parts, so that contracts can be easily generated from common (smaller contract) components. A library of these components could be publicly available, and publicly pre-approved. This would greatly increase the efficiency of measuring trust (transactions in contracts), and greatly increase the ease with which contracts can be created.

Contract-groups built from these pre-approved libraries would achieve consensus with optimal time and resource use. The library increasingly becomes the ONE common good contract. It remains dynamic, constantly re-organising around consensus in every context. Participation within the ecosystem \lq programs\rq\ the contract stack and evolves the library, while the library is used to build contracts (individuals and groups) in consensus. \\

So; \\

\noindent HOC\textsuperscript{1} / HOC\textsuperscript{2} / HOC\textsuperscript{3} is the same path as; \\
 HOC\textsuperscript{3} / HOC\textsuperscript{2} / HOC\textsuperscript{1} is the same path as; \\
  HOC\textsuperscript{1} / HOC\textsuperscript{3} / HOC\textsuperscript{2} etc. \\

\noindent The pre-approved contract component perspective: \\

The network can be viewed from each library contract perspective. For example, the common good might likely agree on a high-level transport contract. That is, how best can all of humanity organise transport from a global perspective, and by what rules? Once approved, why not have many sub-order groups use the transport contract, thereby meeting the common good criteria, and collaborating with all the public for good. This mechanism would operate mechanistically throughout the ecosystem, and scale iteratively to high levels of complexity. \\

\begin{figure}
\centering
\includegraphics[scale=0.7]{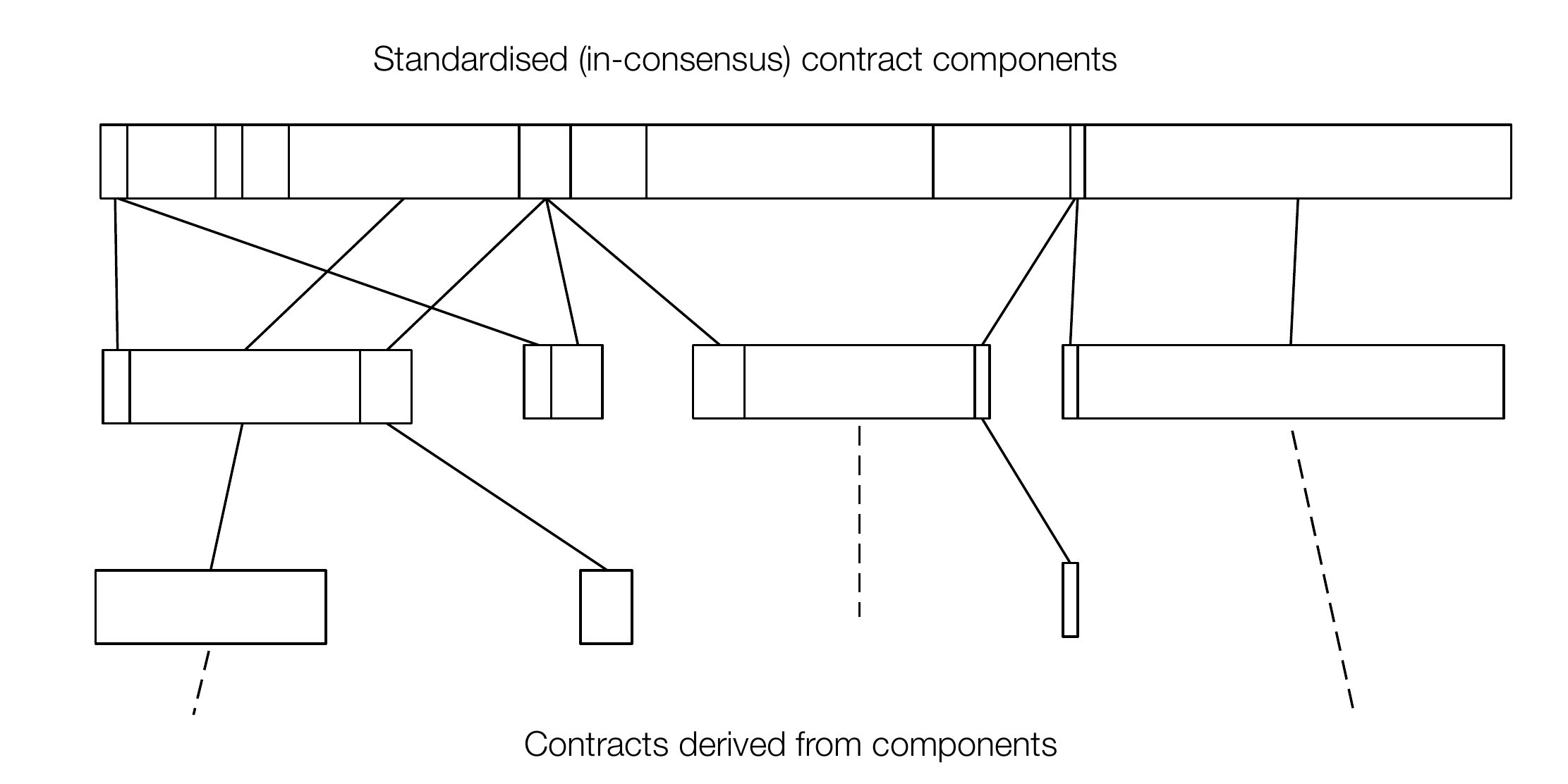}
\caption{Pre-approved and standardised contract components}
\label{components}
\end{figure}

\noindent A difference engine; \\

Looking at Figure \ref{components} on page \pageref{components}, each node/contract/context/purpose is comprised of two parts:

\begin{enumerate}
\item the common component - the \lq position\rq\ not defined locally but within the unified ontology (algorithm), and,
\item the unique component - defined locally and included in the unified ontology (algorithm).
\end{enumerate}

The entire contract stack is formed as a result of the contract differences in the overall ontology.

\subsubsection{Issuance and redemption}

ONE Engine is a fully distributed money issuance computational engine that capitalises industry. All industry that reaches consensus approval by the public, for the common good is fully funded. All members of the public can \lq see\rq\ trusted providers and by contextualised consensus buy products that inherently increase the wealth of the common good.

Labour makes a claim on ONE and is paid with PoAcoin, which exists as a claim. ONE is effectively in debt to labour, until it settles the debt, by redeeming the PoAcoin (a promise) with products in a voluntary exchange. In this two part process, labour is issued PoAcoin (as a claim) and backs its value with free market competitive wage prices, and consumers issue PoAcoin (as redemption) and ONE backs its value with products, at prices that bear all and only their own costs.

When both parties are transacting in the ONE Engine domain (reliably in the consensus state tree) then the free market sets the value of the transaction (and contracts). PoAcoin is the proof of transacting in common purpose, an actual record of the unique transaction the coin issued into at a specific time.

\begin{figure}
\centering
\includegraphics[scale=0.6]{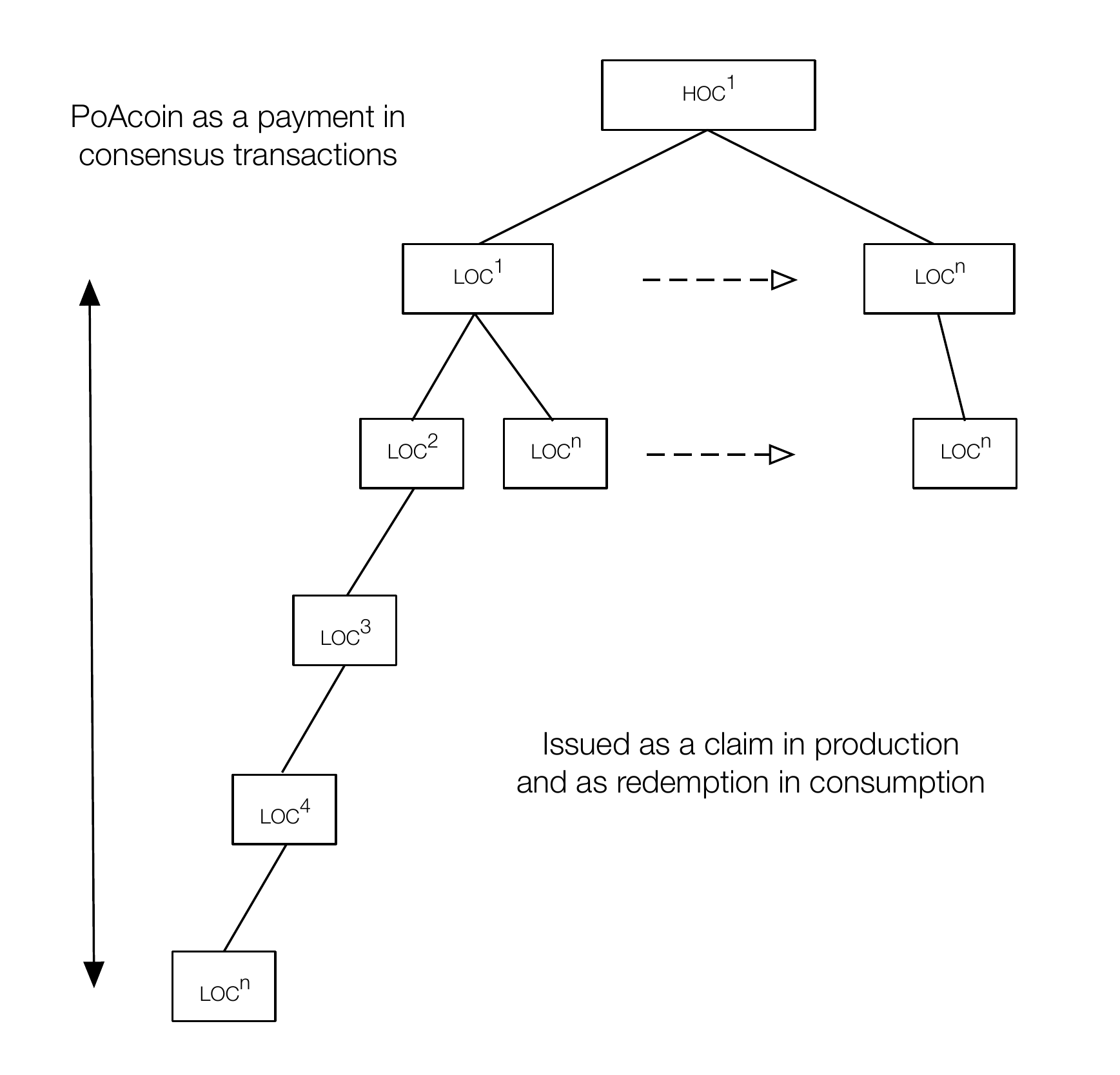}
\caption{Issuance and redemption at the transactional edge}
\label{issuance}
\end{figure}

The ONE Engine contract can \lq see\rq\ any next order contract calling it. Transparency is achieved by making contracts visible to higher and lower order contracts. Trust is defined as a measure (audited) of transactions completed within ONE Engine contract stack.  As this free market organises to optimise for trust and consensus, network wide trust data is trivially easy to see, and model. It becomes obvious that making data relevant (contextualised by contract) and consumable (modelled and visualised effectively) is of high value to everyone. Any contract that meets the consensus requirement of the ONE Engine contract (which is reached increasingly effectively as trust data is generated) is issued PoAcoin. The ONE Engine contract issues PoAcoin into the contract in which the transaction is defined.

PoAcoin is issued at the transactional \lq edge\rq, where goods and services are sold to the contract buying them. Although, typically, issued throughout the contract stack as labour, value added services, tooling and innovation are added to the value chain.

When HOC\textsuperscript{1} has reached consensus approval, it will look to HOC\textsuperscript{2} contracts to fulfil it, which will look to HOC\textsuperscript{3} contracts and so on. Each contract in each chain will require consensus to be approved for PoAcoin issuance. In Figure \ref{issuance} on page \pageref{issuance}, if the contract HOC\textsuperscript{4} needs products and-or services to fulfil its obligations to HOC\textsuperscript{3}, and is approved, PoAcoin is issued into transactions to buy those goods and/or services. From the buyer\vtick s perspective, the onus is on them to find trusted parties and upper order approval, and when this is achieved the buyer is issued money for the transaction, which is spent at the same time. From the seller\vtick s perspective, the onus is on them to serve approved upper order contracts, be transparent and sell approved goods and services. This mechanism can function in great detail, and scale anywhere. \\

\emph{Money is pre-issued as credit down through the stack chain, until it converts and issues, or is redeemed, as money proper at the transactional edge.} \\

As groups are formed in the stack, they can be issued \lq credit\rq\ for budget to complete the group\vtick s purpose. The credit notes issued to each group in a chain simply represent what the current price is for fulfilling the contract. As the group sub-contracts to fulfil its purpose, it would pass on a part of its credit or budget to the next LOC, and so on. These credit notes automatically convert into money, PoAcoin, as the final aspect of the credit note chain reaches its natural conclusion of eventually paying for labour and/or capital goods. At that point, in a final transaction, the ONE engine issues the lowest order aspect of that chain of credit,  money, in a pre-approved (pre-mined) transaction. Ultimately, the credit note is converted to money, and distributed to a person. In this way, no individual or group is \lq given\rq\ money first to then subsequently spend; instead, these credit notes are distributed according to the contract stack, to convert to money when completing a pre-approved transaction. Individuals form groups to collaborate on projects to fulfil on profiting the common good. This is achieved by design-science lead consensus and trust so that, credit follows consensus and trust, and then issues as money only when transactions are completed. In this way there remains an automatic money issuance for industry. \\

\emph{This pre-mines each coin as a contextualised credit note that flows through each chain, until it finally issues. Each coin is a record of its chain path, all the contractual contexts preceding and composing it, both transactional parties and the time of transaction and issuance.} \\

Each coin is a transaction record of the entire production chain-aggregate. Coin data could be analysed and modelled providing requisite data for a genuine computational machine programmed by use, \emph{\lq the world-around accounting system\rq} proposed by R. Buckminster Fuller.

The actual completion of any transaction chain (that \emph{is} the transaction) happens instantaneously through the stack (both \lq up\rq\ and \lq down\rq), at the time of the transaction. Until then, it is represented as a \lq most probable\rq\ (weighted by consensus) contender. \\

Roughly the same process happens, in reverse, in PoAcoin redemption. All PoAcoins in consumer control exist as potential redemption tokens, and at the moment of purchase, become money proper, and transact to redeem the promise made by ONE for the production it has held in trust. On redemption, coins are \lq burned\rq, that is, lose their transactional face value, although still function to provide system wide accounting data. We discuss redemption in more detail throughout the document.

\subsubsection{Consensus}
Given enough accurate global transactional data, fair representation, and the opportunity to be rewarded fairly, the majority of people will be economically incented to use evidence in making decisions in consensus, rather than using guessing, opinion, religion or myth. In other words, the majority of stakeholders in any given context, given the tools, will tend towards agreement based on evidence, particularly when it is seen clearly that this approach produces optimal results.
	
Every transaction \lq votes\rq\ for each relevant contract in the stack chain. Consensus aggregates automatically in stacked contracts. Reasonable use of the system organises the system itself, entirely replacing the idea of government based regulation.

\begin{quotation}
\emph{\lq\lq How I choose to live my life is my vote.\rq\rq}
\end{quotation}

Consensus becomes more probable, simply because, in a system where labour will prefer economic incentives, and the system pays in consensus, people will agree. The market is economically incented to agree, and transact in agreement. And consumers are economically incented to buy consensus products.

\subsubsection{Trust and reputation}

As trust is contextualised throughout the entire economic system, the reputation system organises data in the same unified ontology, creating \lq shared throughout all relevant contexts\rq\ profiles for every one and every thing.

\subsubsection{Auditing}
In order to maintain integrity between the network and the real world, auditing in ONE will become an entire industry, and an important one. Trusted auditors will be well paid for the essential value they provide the economy.

ONE is a tool, designed to optimally represent humanity, and its purpose. It is an abstracted-world-view of our real world. Where, and how these two worlds meet is important. Into every field, in every aspect of this new economy, the industry of auditing would be infused, simply tasked with reporting on how well the on-network view of the world represents the \lq real\rq\ world off-network, and vice versa.

\subsubsection{Externalisation of costs}
By only issuing and redeeming PoAcoin in unique transaction chains, and by structuring all contracts in a unified ontology, all prices reflect all, and only their costs. This optimises for the avoidance of the externalisation of costs.

\subsection{Securing the network}

People, groups, robots and agents will live as contracts on the network, relating within communities defined by context groups.

\subsubsection{Persistant contract images}

\paragraph{Authentication, identity and accounts}

Each on-network account (actor) represents one real off-network person (or \lq thing\rq), and vice versa, and the ONE Engine contract represents, optimally, all people on Earth. \\

The ONE platform offers a radically different method to authenticate participants on the network. Rather than forcing authentication, persistence and authenticity of contracts is economically incented for. This will tend towards an ever increasing defence against Sybil or majority attack scenarios. Persistent contract images contribute to securing the network.

\paragraph{Security, persistent Digital Identity Spectrum image}

Each person, using the network as a tool for organising his or her life, will live as a contract on the network. Each contract will have contextualised public and private aspects, one\vtick s digital identity spectrum (DIS). Each DIS is maintained as an \lq image\rq\ by way of the transparent contracts he or she is in. The ONE mechanism can audit the validity of any DIS as a result of the complex, yet exact, multi-way contracts each one maintains.  Any corruption of a DIS would immediately generate an integrity error. Each DIS comprises an aggregation of all contracts each person is in. In a way, it represents their own personal transaction ledger, as all transactions completed over time in any contract can be seen from a person\vtick s unique perspective. Each person\vtick s community of contractual relationships that co-form one\vtick s DIS is what authenticates each person in a fully contextualised peer way. The more transparent contracts any person maintains on the network, the more persistent and secure their identity becomes.

As the entire ONE contract exists as a unified ontology, that is, is comprised of just one of each type of contract, it is ensured that there be just one of each persons DIS. The \lq identity\rq\ of non-human contracts (things, robots etc) would function in a similar way.

\subsubsection{Proof-of-\emph{X}}

The contracts themselves are the functional equivalent of the block, and transactions are actual transactions within these \lq blocks\rq. However, rather than requiring cryptographic securing of the network using economically arbitrary techniques of cryptographic hashing, the blockcloud allows human use of the mechanism to secure the network, where PoW is actual work transactions, PoS is actual reputation stake and so on. We go on to make distinctions of Proof-of-Identity, Proof-of-Common-Purpose etc.

Rather than being just another blockchain crypto-coin, PoAcoin aims to evolve the blockchain paradigm into what we call \lq the blockcloud\rq, a new kind of very-many-chained blockchain. Each block, or contract, is sustained in consensus state, transacted in by proof-of-work as labour, and each block is unique in a unified ontology of blocks. This becomes a fully contextualised, unified 'cloud' of blocks. Using mechanism and contextualisation, we propose the functional equivalent of a number of \lq proofs\rq.

It is the economically non-arbitrary PoW (actual labour work), PoS (actual reputation) that secures the network in a non-arbitrary way, therefore, ultimately, making it valuable and meaningful economically. This allows money to be issued and redeemed meaningfully, and for all labour, production and consumption to be organised successfully.

Fundamentally, economically arbitrary PoW, PoS or Po\lq X\rq\ systems will always, inherently, fall short of meeting both technical and economic criteria, and, money is far too important a tool to allow for arbitrariness of any kind.

\begin{description}
\item[Proof-of-Work] is actual labour, and is key to the creation and allocation of PoAcoin.
\item[Proof-of-Stake] is performed by economic participants reliably performing transactions in contextualised contexts / contracts over time. People naturally allocate more trust to actions performed reliably over more time. Thus, just as in current cryptographic PoS systems, where coins \lq age\rq\, the trust aspect of our mechanism \lq ages\rq\ and becomes more valuable, naturally.
\item[Proof-of-Identity] (Proof-of-Persistence) A real person may wish to trade anonymously as another identity or contract on the network. People are economically incented to ultimately identify themselves authentically on the network. Reputation quickly becomes one\vtick s most valuable asset, and to allocate trust gained from work as labour to another identity becomes prohibitively expensive. Software robots, or autonomous agents can earn trust, and money, on the network, however, they will not be confused with real people doing labour, and ultimately the profit from robots will naturally be allocated to relevant stakeholders. This forms the foundation of the Autonomy of Things, and the digital identity spectrum and authentication system.
\item[Proof-of-Redemption] (Proof-of-Burn) PoAcoin is redeemed in consumer transactions, fulfilling the life-cycle of money proper.
\item[Proof-of-Common-Purpose] PoCP is derived from the integrated nature of the ONE economy. PoAcoin is only issued and redeemed in common purpose.
\item[Proof-of-Autonomy] The PoAcoin system mechanistically integrates many proofs, into what we call Proof-of-Autonomy.
\end{description}

\section{The Governor}

\subsection{An overview of autonomous free market dynamics}

Certain aspects of an autonomous economy must be automatically governed, as they cannot, and should not be regulated politically or by special interest groups. Automatically regulating certain aspects of the mechanism is key to providing freedom to individuals and the market. \\

All participants choose to participate in specific production, leisure, or consumption contexts at the free market\vtick s rates associated with those contexts. \\

Contextualised representative leadership in each group is organised based on consensus and reputation. \\

\emph{\lq\lq Concentrated power \emph{(force)} is not rendered harmless by the good intentions of those who create it.\rq\rq} \footnote{Milton Friedman} \\

The ONE governor operates two restorative, negative feedback, systems acting in two specific directions. \\

These serve to incent participants to:

\begin{enumerate}
\item Collaborate in consensus
\item Act reputably and transparently
\item Work more or less as required by the whole system
\item Leisure more or less as required by the whole system
\end{enumerate}

\subsubsection{Incentives to labour and leisure}

\begin{figure}
\centering
\includegraphics[scale=0.9]{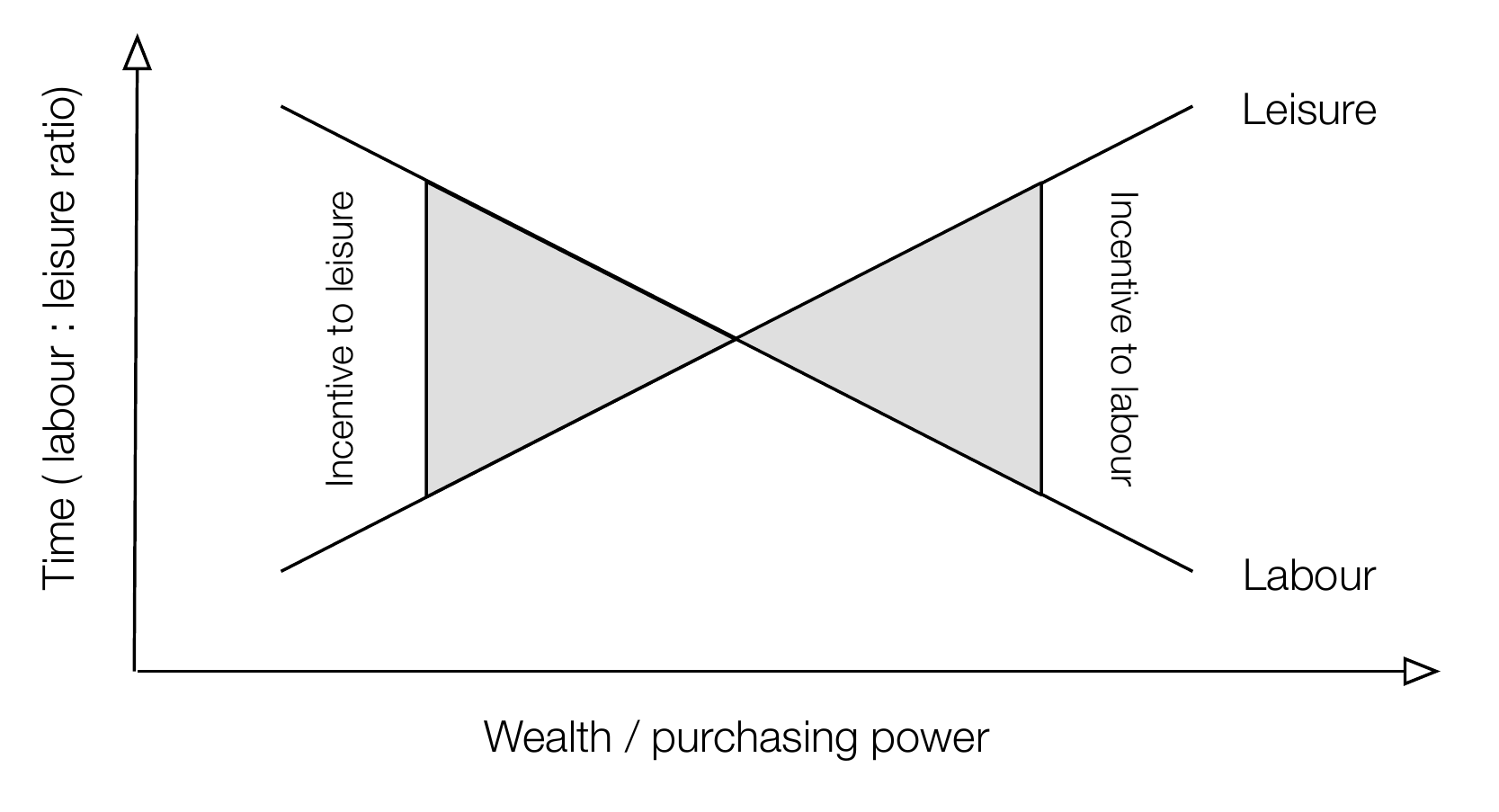}
\caption{Incentives to labour and leisure}
\label{governor}
\end{figure}

Looking at Figure \ref{governor} on page \pageref{governor}, plotting wealth or purchasing power against voluntary time spent either in labour (earning) or leisure (spending), we observe;

\begin{enumerate}

\item As an individual becomes increasingly wealthy, it is probable that time will be spent in leisure rather than labour, to utilise purchasing power. As labour is increasingly \lq under\rq\ supplied, wealth will decrease but wage rates will increase and thus leisure will be incented back to labour. Less labour will produce less products, increasing prices, and decreasing purchasing power also incenting leisure back to labour. \\

And vice versa; \\

\item As an individual becomes decreasingly wealthy, it is probable that time will be spent in labour rather than leisure, to increase purchasing power. As labour is increasingly \lq over\rq\ supplied, wealth will increase but wage rates will decrease and thus labour will be incented back to leisure. More labour will produce more products, decreasing prices, and increasing purchasing power also incenting labour back to leisure.

\end{enumerate}

\subsubsection{Saving}
Saving money (voluntarily not consuming) flattens the leisure curve, and saving time (voluntarily not labouring) flattens the labour curve. The governor still functions in both cases, yet at different rates.

\subsubsection{Purchasing power}

Issuing money for and to labour, and redeeming money for products of labour naturally maintains the purchasing power of the PoAcoin.

\subsubsection{Quality of money - securitisation of value}

Low quality money can\vtick t afford trust, privacy or justice. \\

\emph{\lq\lq The value of money, as we have seen, depends upon the subjective valuations of the people who hold it. And those valuations do not depend solely on the quantity of it that each person holds. They depend also on the quality of the money.\rq\rq} \cite[page 151]{hazlitt} \\

PoAcoin maintains its value over a long time and is assured by the nature of the issuance and redemption mechanism. The ONE economy tends towards enriching the common good, and by default the individual good. PoAcoin's inherent quality is maximised by virtue of its design, founded on consensus agreement, trust, and economically non-arbitrary creation, allocation, backing and redemption. \\

As labour decreases, driving wage and product prices up, there is less production capacity to make products. An under supply of labour and products increases prices. The tendency will be to return to labour as our more essential needs are under supplied, decreasing production capacity of less essential needs. \\

And vice versa; \\

As labour increases, driving wage and product prices down, there is more production capacity to make products. An over supply of labour and products decreases prices. The tendency will be to return to leisure as our more essential needs are over supplied, increasing production capacity of less essential needs. \\

As the market will work earlier and more to meet more essential needs, and later and less to meet less essential needs, this mechanism will find a balance between production and consumption of more and less essential products.

\subsubsection{Quantity of money - the end of \emph{x}-flation}

Inflation is the debasement of a currency's value by it being issued for unproductive industry, increasingly failing to back its value, and/or being allocated to non-producers, biasing purchasing power.

In ONE, labour earns money (as a claim) into the economy, consuming spends money (as a redemption) out of the economy. The volume of total money increases or decreases automatically as a function of these two amounts.

Total labour production earns and maintains purchasing power, so that total consumers have enough money to pay total prices. Total prices are derived from total costs, which, ultimately is a total cost of labour.

As the change in money volume is commensurate with real value (free market labour), and with real production (products of free market labour), inflation and deflation are naturally avoided. \\

\emph{\lq\lq All that is necessary is to have a system of creating new money if the price-level tends to fall and unsaleable goods to stack up, and to destroy it if they get scarcer and prices tend to rise. This is quite impossible under the existing banking system, but quite possible under a rational, scientific, and national system, designed in accordance with the physical realities to which the production and consumption of wealth must conform.\rq\rq} \cite[page 158]{soddy} \\

Ultimately this economy is disinflationary, as we all become wealthier, we don\vtick t need to produce or consume as much. This is the goal of a sustainable and efficient economy. As the total demand for goods and services deflates over time, the free market will have a disinflationary total money supply to buy disinflationary total prices, until they reach a consensus stasis of maintenance and innovation. Then the quality of money is maximised, and the quantity of money stabilises to a merely useful level, yet remains natively dynamic. \\

\emph{\lq\lq Inflation means the creation of money units without commensurate creation of wealth.\rq\rq} \cite[page 32]{riegel} \\

Overall, as individuals become wealthier, and want or need to work less, the rate of increasing wealth will slow. This is good. ONE requires less labour as it becomes more efficient and achieves its goals. Wealthier people will enjoy more leisure and more purchasing power to consume. As people work less, and spend more, less money is issued, and more is burned, thus the over all currency supply decreases, and in dynamic response to these movements. As currency supply decreases, and wealth begins to drop, contributions to the common good (ONE) also slows and cost of living will increase. Thus overall incentive to work again will rise.

The ONE mechanism has all coin accounting data for money issued, and for money redeemed. Money issuance and allocation follows a normal probability distribution.

\subsubsection{Prices}

If there is an \lq over\rq\ supply of labour, supply and demand in free market wages will decrease wage rates, and subsequent prices, however, more labour, and less consumption will tend to increase overall money supply, which will push prices up. \\

And vice versa; \\

\noindent If there is an \lq under\rq\ supply of labour, supply and demand in free market wages will increase wage rates, and subsequent prices, however, less labour, and more consumption will tend to decrease overall money supply, which will pull prices down. \\

This mechanism, will tend to maintain steady price levels, that are commensurate with free market production and consumption.

\subsubsection{Efficiency -  competition, scarcity and monopoly}

There are two distinct types of competition. We shall call them useful or constructive competition, and non-useful or destructive competition. Constructive competition comprises labour competing for wage prices, and consumers competing for product prices. This simple mechanism maintains \lq a level playing field\rq\ or \lq keeps things honest\rq. It\vtick s a way of letting the group of all producers set wage rates, and consumers \lq clear\rq\ the marketplace of products across all contexts and trust ratings. It\vtick s natural and effective. \\

In contrast, having similar producers making similar things compete with each other is destructive competition, and is inherently inefficient, as it pitches common purposes against one another. \\

\emph{\lq\lq Therefore, though I wish to see abundance in everything else, it is in my interest for scarcity to exist in the very thing that it is my business to supply. The greater the scarcity, compared to everything else, in this one thing that I supply, the higher will be the reward that I can get for my efforts.\rq\rq} \cite[page 179]{hazlitt} \\

By unifying each contract, and therefore each producer, there is \lq scarcity\rq\ in the thing produced relative to everything else so that the efficacy of the specialisation of labour is maximised.

So each production centre, or group becomes maximally unique, and maximally \lq scarce\rq\, and therefore maximally valuable relative to every other group. The ONE structure re-engineers the ideas of competition, scarcity and monopoly. Technically, the mechanism keeps competition constructively at the free market edge, while maximising the efficiency of the specialisation of labour through scarcity and allowing each group to be a unique specialist without having monopolistic force, and within system-wide, democratic, consensus governance. \\

\noindent Over-all market efficiency is further enhanced as follows; \\

Money isn\vtick t so much a \lq thing\rq, it\vtick s a process. It is issued as a credit note, a promise to pay, and is redeemed for products later. Without its use in this complete process, \lq money\rq\ as such doesn\vtick t exist. It\vtick s a tool, that, when used enables people to collaborate to build products everyone can use, and enables all costs of production labour to be distributed to all consumers.

When PoAcoin is redeemed, by being exchanged for products, it settles the claim throughout the production chain. If there is a difference in the amount the product cost and how much it is redeemed for, the price paid, then; \\

If the market won\vtick t pay the cost-price, and products are sold at below cost, total money supply increases, tending to increase prices, correcting for the inefficiency of making products that the market isn\vtick t willing to pay for. When there is a subsequent decrease in labour willing to work to make the product, the work could only be done by raising the price of labour, further driving up the cost of the product, for which there is below market value demand. This further dis-incents its production. In this way, the market\vtick s willingness to pay for products, at market costs-prices, drives natural labour demand for each product\vtick s production. \\

That is; unredeemed coins indicate an inefficiency in the market, and will naturally call for a redistribution of labour and production. This has the effect of creating, modifying and dissolving contracts dynamically. \\

\section{Effects}

Our goal is to provide a mechanism that is enabled by free market activity to dynamically balance factors such as money supply, stable prices, purchasing power, wealth, incentives to work, incentives to leisure, and overall employment.

Incenting profitable behaviour and disincenting plunderous behaviour is one goal of a sustainable economy. The only way to \lq get rich\rq\ individually in ONE is to benefit fairly through participating in consensus to increase the wealth of the common good.

\begin{figure}
\includegraphics[scale=0.3]{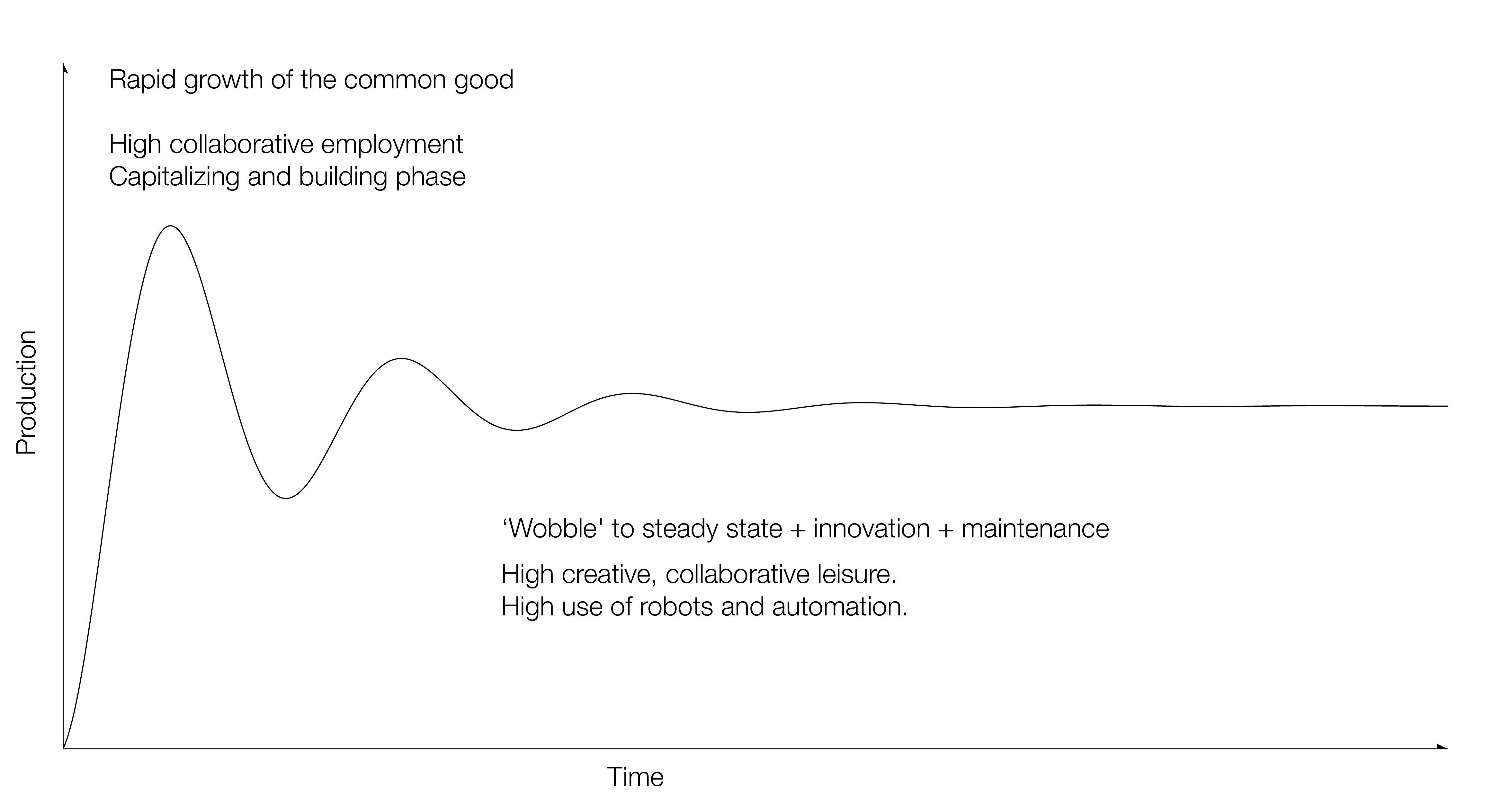}
\caption{Towards a sustainable economy}
\end{figure}

Although the ONE economy will produce products for individual use, it will bias production for the common good, as PoAcoin pays in consensus agreement. As the ratio of provision for the common good to individual good increases, the general cost of living decreases, benefiting all participants. An economy that produces the valuable, rather than being arbitrarily busy allows for eventual true economic stability, and the end of a senseless \lq growth\rq\ economy. It offers an end to the waste and loss of human and Earth\vtick s resources, until humanity, and the earth are wealthy and operating in a dynamic economic balance. \\

\emph{\lq\lq The art of economics consists in looking not merely at the immediate but at the longer effects of any act or policy; it consists in tracing the consequences of that policy not merely for one group but for all groups.\rq\rq} \cite[page 5]{hazlitt}

\subsection{What is going to be different?}

\subsubsection{Generally and socially}

Initially, there would be high collaborative employment, as ONE funds industry to grow the common good, in a building phase, until global society reaches a consensus level of general, and individual well-being. Then the economy would \lq wobble\rq\ towards a stable and sustainable level of production and consumption.

Efficiency and organisation will outperform the need to externalise costs. This creates quality in preference to quantity. Or more specifically, a higher quantity of the \lq higher\rq\ quality, and a lower quantity of the \lq lower\rq\ quality things.

The current model disproportionately centralises wealth to company owners, by aggregating \lq profits\rq\ into corporations. In the ONE economy, money is distributed fairly to all stakeholders automatically. The \lq wealthiest\rq\ (by a relatively small margin) will be those that are trusted most to lead, to take the most responsibility, offer the most value. The \lq poorest\rq\ (by a relatively small margin) will be those who only wish to work enough to get by, and enjoy more leisure. However, the distribution of wealth will be approximately normal in the mathematical sense, ending the \lq wealth gap\rq. In other words, it's fair, and it will end class-ism. The ONE mechanism generates the right amount of high quality money, distributed intelligently.

The idea of nation states and countries may change significantly over time in the ONE system. In the context \lq geography\rq\, the ONE platform will enable incredibly efficient management of resources and activity contextualised by geographical area, so that each place on Earth would participate intelligently and uniquely given its unique features and resources. People may move to the place that is central to a common purpose. In fact, the economic AI, might pay to move people around the Earth as required for optimal common good.

As real world effects of this mechanism enrich the common and individual good, it will become decreasingly important to measure economic success in terms of jobs, employment, GDP and so on. As this economy, and those participating in it become ever increasingly wealthy, it is obvious and right that employment should drop as low as possible, increasing leisure and the use of hardware and software robots. The economy will tend towards sustainable levels of maintenance, plus consensus-approved innovation. Waste, nuisance and harm will be minimised. This will ensure the restoration and safety of the environment and well-being of other species. \\

\subsubsection{Specifically and individually}

The ONE system will (at least);

\begin{enumerate}
\item Decrease the \emph{need} to work, and increase the \emph{freedom} to work.
\item Lower the cost of living.
\item Enable everyone to do what they love with like-minded people.
\item Enable everyone to become \lq wealthy\rq.
\item Enable everyone to live in a rich, safe, sustainable culture.
\end{enumerate}

\lq\lq \emph{We should do away with the absolutely specious notion that everybody has to earn a living. It is a fact today that one in ten thousand of us can make a technological breakthrough capable of supporting all the rest. The youth of today are absolutely right in recognising this nonsense of earning a living. We keep inventing jobs because of this false idea that everybody has to be employed at some kind of drudgery because, according to Malthusian Darwinian theory he must justify his right to exist. So we have inspectors of inspectors and people making instruments for inspectors to inspect inspectors. The true business of people should be to go back to school and think about whatever it was they were thinking about before somebody came along and told them they had to earn a living.}\rq\rq \footnote{R. Buckminster Fuller}

\section{An autonomous global society}

\subsection{Meaningful artificial intelligence}

Current artificial intelligence research focuses on mimicking the brain, using quantum or super computing, and/or \lq neural\rq\ networks. We assert that a meaningful AI must be programmed by global human economic activity. The ONE network allows for the \lq every-person in every-moment\rq\ programming of the system.

The ONE Engine is an algorithmic money issuance machine, forming a contract stack as a continually changing algorithm, and each coin issues through the algorithm in specific contexts and times. The algorithm is \lq programmed\rq\ by human consensus and trust. In this sense, humanity\vtick s use of the mechanism fulfils the role of the \lq oracle\rq\ for a meaningful AI.

The blockcloud is a \emph{\lq process fractal\rq} \footnote{A term coined by Thomas Campbell} within which a meaningful and rich information set is operating. The operator, or oracle, is all people processing it. It probabilistically organises a whole and unified free market. Consensus, trust and money comprise an information set that\vtick s fundamentally useful to human society.

This creates a globally unified body of shared knowledge, transforming the way humanity benefits from research and science.

The expanded, unified ONE contract becomes the memory of the AI. Hardware devices will be neutral viewers onto the network.

\subsection{\emph{Any-context} system-wide graphs}

The \lq social graph\rq\ is only a small part of the data available in the ONE network. The \emph{any-context} graph becomes trivially available, so that graphs from the education, individual, economic, production and consumption contexts, for example, provide rich, economically and socially meaningful, data.

\subsection{\emph{Big} data}

An ontology represents knowledge as a set of concepts within a domain and the relationships between those concepts. \\

\emph{\lq\lq Knowledge representation and knowledge engineering are central to AI research. Many of the problems machines are expected to solve will require extensive knowledge about the world. Among the things that AI needs to represent are: objects, properties, categories and relations between objects - situations, events, states and time; causes and effects - knowledge about knowledge (what we know about what other people know) and many other, less well researched domains. A representation of \lq\lq what exists\rq\rq\ is an ontology: the set of objects, relations, concepts and so on that the machine knows about. The most general are called upper ontologies, which attempt to provide a foundation for all other knowledge.\rq\rq} \footnote{Wikipedia} \\

Currently, \lq big data\rq\ is owned and controlled by corporations. Non-voluntary third party use of personal information creates low quality data under the illusion that it can be made valuable through complex modelling and then sold back to its rightful owners under duress.

The data available within the ONE ecosystem will be rich and meaningful, as trust enables high quality data. Information shared in trust is far more honest, valuable, and meaningful than data shared in not-trust. The ONE mechanism intrinsically generates a global, structured, rich data set. All participants are in voluntary control of personal data and contracts.  

The future is not \lq big data\rq. The future is big trust and collaboration.

\subsection{Web 3.0 and the end of \lq search\rq}

An upper level ontology of common purpose infused with a global economic system will allow for the foundation of a new kind of Web. We believe that it is the ONE structure that will enable a functional Web 3.0.

The economically organised data-set means that any contract (individual or group) can \lq look out\rq\ through its related contracts to the whole system. This enables anyone to see a real-time view of his or her own contract ecosystem, or from any other context. Therefore, search is replaced with \lq contextualised view\rq. By changing the order of filtering, one can explore increasing contextualisation. This replaces search as we know it with intelligent findability. \\

\subsection{The Autonomy of Things}

The Internet of Things is a transitional idea, that must be evolved to create a platform that operates intelligently, and safely. To do this sufficiently well, it must be built on an intelligent economic mechanism. The ONE platform forms the foundation of a new economic operating system, on which \lq things\rq\ can operate.

Software and hardware robots, as contracts on the network, are subject to the same behavioural economics as people. They will come in and out of economic existence, by becoming more or less trustable (transacting in consensus agreement) through their use to the free market.

Any person, thing, device or robot (hardware and software) will have a corresponding contract identity on the network. These would transact within contracts trivially fulfilling reporting and maintenance requirements etc.

This is the only way the Internet of things will work without nuisance. We call it the \lq Autonomy of Things\rq.

\subsection{The new labour and leisure}

When ONE is mature, and has a sufficiently large proportion of all people using it, the job market as we now know it will change profoundly. As the cost of living decreases, wages remain high, and production is organised efficiently ... the need to \emph{work to survive} will shift to the \emph{freedom to work} and to be wealthy in a wealthy culture. Labour demographics will increasingly reflect the alignment of enjoyment and skill relevant to each purpose. Intelligent, collaborative leisure will become commonplace.

The ONE economy, rather than plundering by power over labor, and by externalising costs, will group by common purpose, so that the purpose itself is where power is aggregated, and then labor consents to contribute to that purpose, and is rewarded at the free market rate. Individuals in the group maintain position through consensus and reputation. \\

No person or group has \lq power over labor.\rq\ Labor is governed by purpose and the free market, as it should be. All individuals are free to consent to labor roles, and in fact, power is returned to labor, because the free market allows labor to choose freely where it will consent to contribute. Labor then, is free to follow greatest free market rewards, purposes that most suit it, and groups that function most efficiently to reward in trust and wages.

\subsection{The new corporation and profit}

In today\vtick s economy, if you want to hire people to make products, you form a company, and raise capital or debt, and work through business plans and financial budgets, and then hire labour to make things, and sell it all, to pay all the debt back and keep as much profit as the market will bear. \\

By comparison, within the ONE economy the old idea of corporations and companies is entirely replaced with collaborative groups organised within the unified ontology of purpose. These groups seek consensus \lq up\rq\ and \lq down\rq, that is, pursue consensus in HOCs and generate it in LOCs. With consensus they can hire labour knowing that ONE pays, bearing its costs. This funds industry. So instead of getting financial budget, groups seek consensus \lq budget\rq, economically incented, knowing that ONE has unlimited money. That\vtick s what the coin issues as, a free market valued labour transaction in common purpose.

The goal is to provide a sensible way to allow the ONE production economy, to dynamically produce goods and services for both the common good, and consumers, and dynamically balance both money supply and purchasing power.

\subsubsection{Ownership}

The idea that anyone needs to own \lq companies\rq\ will shift entirely. Private or individual ownership remains, but group ownership will be contextually distributed to all group stakeholders. That is, as an individual stakeholder of the global water infrastructure group, you might own a very small, contextualised by where you live, part of all global water infrastructure. Ownership of all infrastructure, is contextually distributed to all people. That is, you would likely own a small part of education, food, transport, housing etc.

While the ONE mechanism can contextually share assets and capital equipment, we retain private and individual ownership of the products of industry, and protection of private property.

\subsubsection{Profit}

In the current economy, producers sell products at cost plus as much margin as the market, and competition will bear,  and that\vtick s called \lq making money\rq. However, there\vtick s no accounting for externalisation of costs, or the fact that it\vtick s been capitalised by debt, and the aggregate global affect of interest. This fragments production, and consumption into arbitrary priorities rather than organising both by consensus and trust, and ultimately value.

The way ONE is structured \lq profit\rq\ does not exist in the traditional sense, that is, in terms of corporate \lq profit\rq. Groups produce wealth for the common good, and yet the group itself doesn't charge the marketplace more than its costs. Group profit in ONE is simply the gain in wealth of the common good. Individual profit is gained through earning wages and increasing reputation. We maintain the \lq profit motive\rq\ to incent labour.

\subsection{The end of monopoly, advertising and competition as we know it}

When very-many companies compete with each other to produce the same products for the same markets, the economy becomes inefficient and fragmented. We define this as non-useful competition. The debt based economy forces success through \lq bigness\rq\, effectively encouraging monopolies, and growth through acquisition. Monopolies \lq grow\rq\ through an unfair advantage over their \lq competition\rq.

In contrast, within the ONE system, \lq companies\rq\ are replaced with unique groups each defined by their purpose. Each unique group doesn't compete directly with any other group, they compete for overall consensus, and seek to earn trust for reputation. This completely redefines the way industry and the marketplace operate. Each group is a \lq monopoly\rq\ , but not beyond its context.

So, we define useful competition as labour competing for free market wages. Highly efficient groups each fulfil unique purposes, and are intelligently findable within the entire economic system. This systemic dynamic resolves as a highly-efficient, and organised production mechanism.

This renders advertising obsolete.

\subsection{The law}

All participants in the ONE economy would have a legal AI at their disposal. \\

\emph{\lq\lq Life, liberty, and property do not exist because men have made laws. On the contrary, it was the fact that life, liberty, and property existed beforehand that caused men to make laws in the first place.\rq\rq} \cite[page 2]{bastiat} \\

The ONE mechanism forms a democratically governed, economically driven, legal and justice system. Directly, by preventing plunder and avoiding the externalisation of costs, and indirectly, by economically incenting everyone to wealth through cooperation and consensus. By making consensus and reputation more valuable than money, people become increasingly self-regulating. \\

\emph{\lq\lq No legal plunder: This is the principle of justice, peace, order, stability, harmony, and logic. Until the day of my death, I shall proclaim this principle with all the force of my lungs.\rq\rq} \cite[page 15]{bastiat} \\

\emph{\lq\lq Peace cannot be kept by force; it can only be achieved by understanding.\rq\rq} \footnote{Albert Einstein}

\subsection{Education}

All participants in the ONE economy would have an education AI at their disposal. \\

The education aspect will exist in every context creating a fully contextualised learning system. This allows for contextualised knowledge and skill sharing throughout the global platform. More specifically; the context \lq education\rq\ will exist within every other context, and automatically will fulfil the requirements of a next generation education platform.

\subsection{News}

All participants in the ONE economy would have a news AI at their disposal. \\

With every transaction in every context being visible to everyone it becomes trivially easy to provide a system-wide service that allows any user or group to filter the world's activities in any contextually useful way. News reflects actual global economic activity, and can be viewed without bias of any special interest group.

\subsection{The stock market}

It has become evident that the stock-market model, is an extremely inefficient way to fund industry and distribute its profits to stakeholders. It is based on economically arbitrary price and trading incentives, and it systematically externalises costs, and functions with near endless arbitrary economic properties.

Ideally, the stock market as we know it now, won\vtick t exist in the ONE Engine framework. Instead of the share model for company ownership, asset ownership and control is contextually distributed throughout each group's stakeholders.

Successful groups will display upward \lq trust-trending\rq\, indicative of an investable group. However, rather than investing capital, the market would invest consensus and trust, which would naturally convert to capital, and higher wages. Successful groups benefit all stakeholders contextually. The reverse would be the case for downward trust-trending groups.

This naturally causes groups to be created, modified, or dissolved according to economic use.

\subsection{Insurance and assurance}

The current insurance industry model makes no sense in the ONE economy. Modern insurance relies on information asymmetry, and pools risk on behalf of its customers. Risk in ONE becomes trivially easy to model, and reparations assurable. Risk could be largely mitigated through design. Why pay for billions of dollars in damages and reparations for car accidents, for example, when we could build a safe, global public transport system that largely replaces cars? The ONE mechanism itself becomes the assurance system.

The ONE model minimises \lq moral hazard\rq\ \footnote{\emph{From Wikipedia} - \lq\lq Moral Hazard : In economics, moral hazard occurs when one person takes more risks because someone else bears the burden of those risks. A moral hazard may occur where the actions of one party may change to the detriment of another after a financial transaction has taken place.

Moral hazard occurs under a type of information asymmetry where the risk-taking party to a transaction knows more about its intentions than the party paying the consequences of the risk. More broadly, moral hazard occurs when the party with more information about its actions or intentions has a tendency or incentive to behave inappropriately from the perspective of the party with less information.\rq\rq} by having all transactional data available and organised, and by not externalising costs.

\subsection{Health}

All participants in the ONE economy would have a health AI at their disposal. \\

Healthcare is a fundamental aspect of the common good. Global consensus would likely dictate that an effective healthcare system would be available, and ONE would fund it as required. No-one would be excluded, or would need health insurance. The ONE healthcare system would attract the most reputable labour, and organise a world-scale health care system like we've never seen before.

\subsection{The social network}

The ONE mechanism naturally causes a next-generation \lq social network\rq. Although, by its nature it is significantly more meaningful, and functional than the social networks of today. Not only is privacy inherently protected, technically and economically, but the social aspect of the ONE network is simply an aspect of the full ecosystem. The contextualised views, and inherently meaningful economic nature of this system would be vastly superior than current  fragmented, corporate owned and controlled platforms.

\subsection{Gaming, forecasting and modelling}

Running as an API at the edge, any aspect of the computational network can be given a \lq view\rq\ through an app. These apps will be designed to provide a game type scenario for modelling aspects of the network using global real-world data. \lq Winning\rq\ game scenarios (those that solve challenges, or better fulfil contracts) can be seen by the ONE mechanism and adopted automatically. This allows a gaming type ecosystem to model mankind\vtick s future in real-time. It becomes the innovative edge of the marketplace, and assigns new champions active, respected, and well-paid roles in the marketplace. Gaming, or modelling using the global economy data-set would be available through the system AI, that is, every context could have a gaming or modelling aspect.

\section{Discussion}

\subsection{The transition}

\subsubsection{Person by person in every context}

We expect interest and investment in superior systems to grow. When ONE is operational, we would expect a transition from one system to the next to occur reasonably naturally, person by person and group by group, as they become interested in a new economy.

\section{Conclusion}

We believe that Autonomics is the science of human evolution. Without optimising money, governance and reputation systems, we, as a species, are severely limited in our ability to produce, organise, relate and expand. Without an efficient and effective money tool, humanity can\vtick t optimise for survival.

\subsection{Next steps}

To prove the function and efficacy of the system, and to build a stable and un-exploitable platform, we suggest;

\subsubsection{Agent-based computational economic modelling}

We suggest agent based modelling and \lq dynamic program analysis\rq\ \footnote{\emph{From Wikipedia} - \lq\lq Dynamic program analysis is the analysis of computer software that is performed by executing programs on a real or virtual processor. For dynamic program analysis to be effective, the target program must be executed with sufficient test inputs to produce interesting behaviour.\rq\rq} to ensure the system be built correctly and suitably analysed.

\subsubsection{Technical proof of concept}

Following a modelling phase, we\vtick d expect a technical build followed by a test deployment, probably as a country specific \lq mine\rq. This would start PoAcoin in one country in the \lq geography\rq\ aspect of the network.

\subsubsection{Challenges}

As an \lq open source\rq\ project, developed by very many people collaborating together, we expect to see the ONE platform reach significant and sustainable development. We believe the inherent benefits that this economy offers, will draw sufficient numbers to develop a project of this scope and vision.

\newpage

\bibliographystyle{plain}
\bibliography{autonomics}

\end{document}